\documentclass[11pt]{article}
\usepackage{graphicx}
\usepackage{fullpage}
\usepackage{amsmath}
\usepackage{float}

\usepackage[authoryear]{natbib}

\usepackage{amsfonts}
\usepackage{amssymb}
\usepackage{amsthm}
\usepackage{graphicx,shortvrb}
\usepackage{color}
\usepackage[OT1]{fontenc}

\usepackage{charter}

\usepackage{hyperref} 
\newcommand{\rref}[1]{\hyperref[#1]{\ref*{#1}}}
\hypersetup{colorlinks=true, citecolor=blue, linkcolor=blue}

\newcommand{\ale}[1]{\textcolor{red}{{\bf ALE $\star$ #1 $\star$}}}

\newenvironment{remark}[1][Remarks]{\begin{trivlist}
\item[\hskip \labelsep {\bfseries #1}]}{\end{trivlist}}

\newtheorem{theorem}{Theorem}[section]
\newtheorem{lemma}[theorem]{Lemma}
\newtheorem{proposition}[theorem]{Proposition}
\newtheorem{corollary}[theorem]{Corollary}

\newtheorem{question}[theorem]{Question}

\theoremstyle{definition}

\newtheorem{example}[theorem]{Example}

\newcommand{\m}[1]{\mathrm{#1}}

\title{Maximum Likelihood Estimation in Latent Class Models For  Contingency Table Data}
\author{Stephen E. Fienberg\\
Department of Statistics, Machine\\ Learning Department and Cylab\\
Carnegie Mellon University\\
Pittsburgh, PA 15213-3890 USA
\and
Patricia Hersh\\
Department of Mathematics\\
Indiana University\\
Bloomington, IN 47405-7000 USA
\and
Alessandro Rinaldo\\
Department of Statistics\\
Carnegie Mellon University\\
Pittsburgh, PA 15213-3890 USA  
\and
Yi Zhou\\
Machine Learning Department\\
Carnegie Mellon University\\
Pittsburgh, PA 15213-3890 USA}
\date{}

\begin{document}
\maketitle

\begin{abstract}
Statistical models with latent structure have a history going back to the 1950s and have seen widespread use in the social sciences and, more recently, in computational biology  and in machine learning. Here we study the basic latent class model proposed originally by the sociologist Paul F. Lazarfeld for categorical variables, and we explain its geometric structure. We draw parallels between the statistical and geometric properties of latent class models and we illustrate geometrically the causes of many problems associated with maximum likelihood estimation and related statistical inference. In particular, we focus on issues of non-identifiability and determination of the model dimension, of maximization of the likelihood function and on the effect of symmetric data.  We illustrate these phenomena with a variety of synthetic and real-life tables, of different dimension and complexity. Much of the motivation for this work stems from the ``100 Swiss Francs" problem, which we introduce and describe in detail.
\end{abstract}

\pagebreak

\tableofcontents



\pagebreak

\section{Introduction}

Latent class (LC) or latent structure analysis models were introduced in the 1950s in the social science literature to model the distribution of dichotomous attributes based on a survey sample from a populations of individuals organized into distinct homogeneous classes on the basis of an unobservable attitudinal feature. See \cite{ANDERSON:54}, \cite{GIBSON:55}, \cite{MAD:60} and, in particular, \cite{LS:68}. These models were later generalized in \cite{GOODMAN:74}, \cite{HAB:74}, \cite{CLOGGOODMAN:84} as models for the joint marginal distribution of a set of manifest categorical variables, assumed to be conditionally independent given an unobservable or latent categorical variable, building upon the then recently developed literature on log-linear models for contingency tables. 
More recently, latent class models have been described and studied  as a special cases of a larger class of directed acyclic graphical models with hidden nodes, sometimes referred to as Bayes nets, Bayesian networks, or causal models, e.g., see \cite{LAU:96}, \cite{PRONET:99}, \cite{HT:03} and, in particular, \cite{GHKM:01}. A number of recent papers have established fundamental connections between the statistical properties of latent class models and their algebraic and geometric features, e.g., see  \cite{SETSMITH:98,SETSMITH:05}, \cite{CROFTSMITH:03}, \cite{RG:05},\cite{WATA:01} and \cite{BAYESNET:04}.

Despite these recent important theoretical advances, the basic statistical tasks of estimation, hypothesis testing and model selection remain surprisingly difficult and, in some cases, infeasible tasks, even for small latent class models. Nonetheless, LC models are widely used and there is a ``folklore" associated with estimation in various computer packages implementing algorithms such as EM for estimation purposes, e.g., see~\cite{UEBERSAX:06a,UEBERSAX:06b}.

The goal of this article is two-fold. First, we offer a simplified geometric and algebraic description of LC models and draw parallels between their statistical and geometric properties. The geometric framework enjoys notable advantages over the traditional statistical representation and, in particular, offers natural ways of representing singularities and non-identifiability problems. Furthermore, we argue that the many statistical issues encountered in fitting and interpreting LC models are a reflection of complex  geometric attributes of the associated set of probability distributions. 
Second, we illustrate with examples, most of which quite small and seemingly trivial, some of the computational, statistical and geometric challenges that LC models pose.
In particular, we focus on issues of non-identifiability and determination of the model dimension, of maximization of the likelihood function and on the effect of symmetric data. We also show how to use symbolic software from computational algebra to obtain a more convenient and simpler parametrization and for unravelling the geometric features of LC models. These strategies and methods should carry over to more complex latent structure models, such as in \cite{BMZR:97}.

In the next section, we describe the basic latent class model and introduce its statistical properties and issues, and we follow that, in Section 3, with a discussion of the geometry of the models. In Section 4, we turn to our examples exemplifying identifiability issues and the complexity of the likelihood function, with a novel focus on the problems arising from symmetries in the data. Finally, we present some computational results for two real-life examples, of small and very large dimension, and remark on the occurrence of singularities in the observed Fisher information matrix.

\section{Latent Class Models for Contingency Tables}
Consider $k$ categorical variables, $X_1, \ldots, X_k$, where each $X_i$ takes value on the finite set $[d_i] \equiv \{ 1, \ldots, d_i \}$. Letting  $\mathcal{D} = \bigotimes_{i=1}^k [d_i]$, $\mathbb{R}^\mathcal{D}$ is the vector space of of $k$-dimensional arrays of the format $d_1 \times \ldots \times d_k$, with a total of $d = \prod_i d_i$ entries. 
The cross-classification of $N$ independent and identically distributed realizations of  $(X_1, \ldots, X_k)$ produces a random integer-valued vector ${\bf n} \in \mathbb{R}^\mathcal{D}$, whose coordinate entry 
$
{\bf n}_{i_i, \ldots, i_k}
$
corresponds to the number of times the label combination $(i_1, \ldots, i_k)$ was observed in the sample, for each $(i_1, \ldots, i_k) \in \mathcal{D}$. The  table ${\bf n}$ has a $\m{Multinomial}_d (N, {\bf p})$ distribution, where ${\bf p}$ is a point in the $(d-1)$-dimensional probability simplex $\Delta_{d-1}$ with coordinates
\[
p_{i_1, \ldots, i_k} = Pr\left\{ (X_1, \ldots, X_k) = (i_1, \ldots, i_k)  \right\}, \quad \quad (i_1, \ldots, i_k) \in \mathcal{D}.
\]

Let $H$ be an unobservable latent variable, defined on the set $[r] = \{ 1, \ldots, r\}$. In its most basic version, also known as the \textit{naive Bayes model}, the LC  model postulates that, conditional on $H$, the variables $X_1, \ldots, X_k$ are mutually independent. Specifically, the joint distributions of $X_1, \ldots, X_k$ and $H$ form the subset $\mathcal{V}$ of the probability simplex $\Delta_{dr -1}$ consisting of points with coordinates
\begin{equation}\label{eq:condind}
p_{i_1, \ldots, i_k, h} = p_1^{(h)}(i_1) \ldots  p_k^{(h)}(i_k) \lambda_h, \quad \quad (i_1, \ldots, i_k, h) \in \mathcal{D} \times [r],
\end{equation}
where $\lambda_h$ is the marginal probability $Pr\{ H = h\}$ and  $p_l^{(h)}(i_l)$ is the conditional marginal probability $Pr \{ X_{l} = i_l | H = h \}$, which we assume to be strictly positive for each $h \in [r]$ and $(i_1, \ldots, i_k) \in \mathcal{D}$. 

The log-linear model specified by the polynomial mapping (\ref{eq:condind}) is a decomposable graphical model \citep[see, e.g,][]{LAU:96} and $\mathcal{V}$ is the image set of a homeomorphism from the parameter space
\[
\begin{array}{rcl}
\Theta & \equiv & \left\{\theta \colon \theta =  (  p_1^{(h)}(i_1) \ldots  p_k^{(h)}(i_k), \lambda_h), (i_1, \ldots, i_k, h) \in \mathcal{D} \times [r] \right\}\\
& = & \bigotimes_i \Delta_{d_i - 1} \times \Delta_{r-1},\\
\end{array}
\]
so that global identifiability is guaranteed. The remarkable statistical properties of this type of model and the geometric features of the set $\mathcal{V}$ are well understood.
Statistically, equation (\ref{eq:condind}) defines a linear exponential family of distributions, though not in its natural parametrization. The maximum likelihood estimates, or MLEs, of $\lambda_h$ and $p_l^{(h)}(i_l)$ exist if and only if the minimal sufficient statistics, i.e., the empirical joint distributions of $(X_i,H)$ for $i=1,2,\ldots,k$, are strictly positive and are given in closed form as rational functions of the observed two-way marginal distributions between  $X_i$ and  $H$ for $i=1,2,\ldots,k$. The log-likelihood function is strictly concave and the global maximum is always attainable, possibly on the boundary of the parameter space. Furthermore, the asymptotic theory of  goodness-of-fit testing is fully developed. The statistical problem arises because $H$ is latent and unobservable.

 Geometrically, we can obtain the set $\mathcal{V}$ as the intersection of $\Delta_{dr -1}$ with an affine variety \citep[see, e.g.,][]{CLOS:96} consisting of the solutions set of a system of $r \prod_i {d_i \choose 2}$ homogeneous square-free polynomials. 
For example, when $k=2$, each of these polynomials take the form of quadric equations of the type
\begin{equation}\label{eq:odds}
p_{i_1,i_2,h} p_{i_1',i_2',h} = p_{i_1',i_2,h} p_{i_1,i_2',h},
\end{equation}
with $i_1 \neq i_1'$, $i_2 \neq i_2'$ and for each fixed $h$. Equations of the form (\ref{eq:odds}) are nothing more than conditional odds ratio of $1$ for every pair $(X_i,X_{i^\prime})$ given $H=h$and, for each given $h$, the coordinate projections of the first two coordinates of the points satisfying  (\ref{eq:odds}) trace the surface of independence inside the simplex $\Delta_{d-1}$.
The strictly positive points in $\mathcal{V}$ form a smooth manifold whose dimension is $r \prod_i (d_i -1) + (r-1)$ and  whose co-dimension corresponds to the number of degrees of freedom. The singular points in $\mathcal{V}$ all lie on the boundary of the simplex $\Delta_{dr-1}$ and identify distributions with degenerate probabilities along some coordinates. The singular locus of $\mathcal{V}$ can be described similarly in terms of stratified components of $\mathcal{V}$, whose dimensions and co-dimensions can also be computed explicitly.

Under the LC model, the variable $H$ is unobservable and the new model $\mathcal{H}$ is a $r$-class mixture over the exponential family of distributions prescribing mutual independence among the manifest variables $X_1, \ldots, X_k$. Geometrically, $\mathcal{H}$ is the set of probability vectors in $\Delta_{d-1}$ obtained as the image of the marginalization map from $\Delta_{dr-1}$ onto $\Delta_{d-1}$ which consists of taking the sum over the coordinate corresponding to the latent variable. Formally, $\mathcal{H}$ is made up of of all probability vectors  in $\Delta_{d-1}$ with coordinates satisfying the \textit{accounting equations} \citep[see, e.g.,][]{LS:68}
\begin{equation}\label{eq:margmap}
p_{i_1, \ldots, i_k} = \sum_{h \in [r]} p_{i_1, \ldots, i_k, h} = \sum_{h \in [r]} p_1^{(h)}(i_1) \ldots  p_k^{(h)}(i_k) \lambda_h, 
\end{equation}
where $(i_1, \ldots, i_k, h) \in \mathcal{D} \times [r]$.

Despite being expressible as a convex combination of very well-behaved models, even the simplest form of the LC model (\ref{eq:margmap}) is far from well-behaved and, in fact, shares virtually none of the properties of the standard log-linear models (\ref{eq:condind}) described above. In particular, latent class models described by equations (\ref{eq:margmap}) do not define exponential families, but instead belong to a broader class of models called stratified exponential families \citep[see][]{GHKM:01}, whose properties are much weaker and less well understood. {\it The minimal sufficient statistics for an observed table ${\bf n}$ are the observed counts themselves and we can achieve  no data reduction via sufficiency.} The model may not be identifiable, because for a given ${\bf p} \in \Delta_{d-1}$ defined by (\ref{eq:margmap}), there may be a subset of $\Theta$, known as the \textit{non-identifiable space}, consisting of parameter points all satisfying the same accounting equations. The non-identifiability issue has in turn considerable repercussions for the determination of the correct number of degrees of freedom for assessing model fit  and, more importantly, on the asymptotic properties of standard model selection criteria (e.g. likelihood ratio statistic and other goodness-of-fit criteria such as BIC, AIC, etc), whose applicability and correctness may no longer hold. 

Computationally,   maximizing  the log-likelihood can be a rather laborious and difficult task, particularly for high dimensional tables, due to lack of concavity, the presence of local maxima  and saddle points, and singularities in the observed Fisher information matrix.   Geometrically, $\mathcal{H}$ is no longer a smooth manifold on the relative interior of $\Delta_{d-1}$, with singularities even at probability vectors with strictly positive coordinates, as we show in the next section. The problem of characterizing the singular locus of $\mathcal{H}$ and of computing the dimensions of its stratified components (and of the tangent spaces and tangent cones of its singular points) is of statistical importance: singularity points of  $\mathcal{H}$ are probability distributions of lower complexity, in the sense that they are specified by lower-dimensional subsets of $\Theta$, or, loosely speaking, by less parameters. Because the sample space is discrete, although the singular locus of $\mathcal{H}$  has typically Lebesgue measure zero, there is nonetheless a positive probability that the maximum likelihood estimates end up being either a singular point in the relative interior of the simplex $\Delta_{d-1}$ or a point on the boundary. In both cases, standard asymptotics for hypothesis testing and model selection fall short.

\section{Geometric Description of Latent Class Models}\label{sec:geom}

In this section, we give a geometric representation of latent class models, summarize existing results and point to some of the relevant mathematical literature. For more details, see \cite{BAYESNET:04} and \cite{GARCIA:04}.

The latent class model defined by (\ref{eq:margmap}) can be described as the set of all convex combinations of all $r$-tuple of points lying on the surface of independence inside $\Delta_{d-1}$. Formally, let 
\[
\begin{array}{ccll}
\sigma \colon &  \Delta^{d_1 -1} \times \ldots \times \Delta^{d_k -1} & \rightarrow & \Delta_{d - 1}\\
& (p_{1}(i_1), \ldots, p_{k}(i_k))  & \mapsto & \prod_j p_j(i_j)\\
\end{array}
\]
be the map that sends the vectors of marginal probabilities into the $k$-dimensional array of joint probabilities for the model of complete independence. The set $\mathcal{S} \equiv \sigma(\Delta^{d_1 -1} \times \ldots \times \Delta^{d_k -1})$ is a manifold in $\Delta_{d-1}$ known in statistics as the surface of independence and in algebraic geometry \citep[see, e.g.][]{HARRIS:92} as (the intersection of $\Delta_{d-1}$ with) the Segre embedding of $\mathbb{P}^{d_1 - 1} \times \ldots \times \mathbb{P}^{d_k-1}$ into $\mathbb{P}^{d-1}$. The dimension of $\mathcal{S}$ is $\prod_i (d_i - 1)$, i.e., the dimension of the corresponding decomposable model of mutual independence.
The set $\mathcal{H}$ can then be constructed geometrically as follows. Pick any combination of $r$ points along the hyper-surface $\mathcal{S}$, say ${\bf p}^{(1)}, \ldots, {\bf p}^{(r)}$,  
and determine their convex hull, i.e. the convex subset of $\Delta_{d-1}$ consisting of all points of the form $\sum_{h} {\bf p}^{(h)} \lambda_h$, for some choice of $(\lambda_1, \ldots, \lambda_r) \in \Delta_{r-1}$. The coordinates of any point in this new subset satisfy, by construction, the accounting equations (\ref{eq:margmap}). In fact, the closure of the union of all such convex hulls is precisely the latent class model $\mathcal{H}$. In algebraic geometry, $\mathcal{H}$ would be described as the intersection of $\Delta_{d-1}$ with the $r$-th secant variety of the Segre embedding mentioned above.

\begin{figure}[ht]
	\centering
	\includegraphics[width=4in]{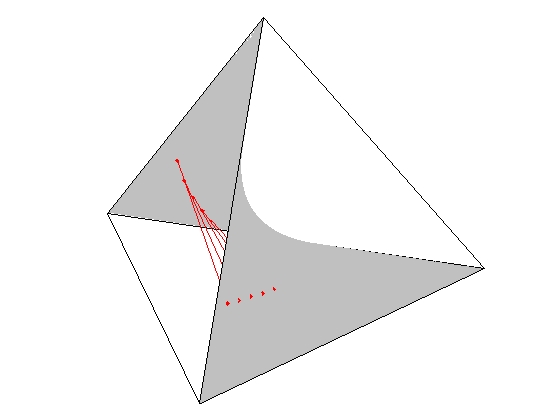}
	\caption{Surface of independence for the $2 \times 2$ table with $3$ secant lines.}
	\label{fig:secant}
\end{figure}

\begin{example}\label{ex:2x2}
The simplest example of a latent class model is for a $2 \times 2$ table with $r=2$ latent classes. The surface of independence, i.e. the intersection of the simplex $\Delta_3$ with the Segre variety, is shown in Figure \ref{fig:secant}. The secant variety for this latent class models is the union of all the \textit{secant lines}, i.e. the lines connecting any two distinct points lying on the surface of independence. Figure \ref{fig:secant} displays three such secant lines. It is not to hard to picture that the union of all such secant lines is the enveloping simplex $\Delta_3$ and, therefore, $\mathcal{H}$ fills up all the available space \citep[for formal arguments, see][Proposition 2.3]{CGG:02}.
\end{example}

The model $\mathcal{H}$, thought of as a portion of the $r$-th secant variety to the Segre embedding, is not a smooth manifold. Instead, it is a semi-algebraic set \cite[see, e.g.,][]{BEN:90}, clearly singular on the boundary of the simplex, but also at strictly positive points along the $(r-1)$st secant variety, (both of Lebesgue measure zero). This means that the model is singular at all points in $\mathcal{H}$ which satisfy the accounting equations with one or more of the $\lambda_h$'s equal to zero. In Example \ref{ex:2x2} above, the surface of independence is a singular locus for the latent class model. From the statistical viewpoint, singular points of $\mathcal{H}$ correspond to simpler models for which the number of latent classes is less than $r$ (possibly 0). As usual, for these points one needs to adjust the number of degrees of freedom to account for the larger tangent space.

Unfortunately, we have no general  closed-form expression for computing the dimension of $\mathcal{H}$ and the existing results only deal with specific cases. 
Simple considerations allow us to compute an upper bound for the dimension of $\mathcal{H}$, as follows.
As Example \ref{ex:2x2} shows, there may be instances for which $\mathcal{H}$ fills up the entire simplex $\Delta_{d-1}$, so that $d-1$ is an attainable upper bound. Counting the number of free parameters in (\ref{eq:margmap}), we can see that this dimension cannot exceed  $r  \sum_i (d_i - 1) + r - 1$, c.f.,  \cite[page 219]{GOODMAN:74}. This number, the \textit{standard dimension}, is the dimension of the fully observable model of conditional  independence. Incidentally, this value can be determined mirroring the geometric construction of $\mathcal{H}$ as follows (c.f.,  \cite{GARCIA:04}). The number $ r \sum_i (d_i - 1)$ arises from the choice of $r$ points along the $\sum_i (d_i - 1)$-dimensional surface of independence, while the term $r-1$ accounts for the number of free parameters for a generic choice of $(\lambda_1, \ldots, \lambda_r) \in \Delta_{r-1}$. Therefore, we conclude that the dimension of $\mathcal{H}$ is bounded by 
\begin{equation}\label{eq:expdim}
\min \left\{ d-1, r  \sum_i (d_i - 1) + r - 1 \right\},
\end{equation}
a value known in algebraic geometry as the \textit{expected dimension} the variety $\mathcal{H}$.  

Cases of latent class models with dimension strictly smaller than the expected dimension have been known for a long time, however.  In the statistical literature, \cite{GOODMAN:74} noticed that the latent class models for 4 binary observable variables and a 3-level latent variable, whose expected dimension is $14$, has dimension $13$. In algebraic geometry, secant varieties with dimension smaller than the expected dimension (\ref{eq:expdim}) are called \textit{deficient} \citep[e.g., see][]{HARRIS:92}. In particular, Exercise 11.26 in \cite{HARRIS:92} gives an example of deficient secant variety, which corresponds to a latent class model for a 2-way table with a latent variable taking on $2$ values. In this case, the deficiency is $2$, as is demonstrated  below in equation (\ref{eq:2way}).
The true or \textit{effective} dimension of a latent class model, i.e. the dimension of the semi-algeraic set $\mathcal{H}$ representing it, is crucial for establishing identifiability and for computing correctly the number of degrees of freedom. In fact, if a model is deficient, then that the pre-image of each probability array in $\mathcal{H}$ arising from the accounting equations is a subset (in fact, a variety) of $\Theta$ called the \textit{non-dentifiable subspace}, with dimension exactly equal to the deficiency itself. Therefore, a deficient model is non-identifiable, with adjusted degrees of freedom equal to number of degrees of freedom for the observable graphical model plus the value of the deficiency.

Theoretically, it is possible to determine the effective dimension of $\mathcal{H}$ by computing the maximal rank of the Jacobian matrix for the polynomial mapping from $\Theta$ into $\mathcal{H}$ given coordinatewise by (\ref{eq:margmap}). In fact, \cite{GHKM:01} showed that this value is equal to the dimension of $\mathcal{H}$ almost everywhere with respect to the Lebsegue measure, provided the Jacobian is evaluated at strictly positive parameter points $\theta$. These symbolic evaluations, however, require the use of symbolic software which can only handle small tables and models, so that, in practice, computing the effective dimension of a latent class model is computationally difficult and often unfeasible.

Recently, in the algebraic-geometry literature, \cite{CGG:02,CGG:03} have obtained explicit formulas for the effective dimensions of some secant varieties which are of statistical interest. In particular, they show that for $k=3$ and $r \leq \min\{ d_1, d_2, d_3 \}$, the latent class model has the expected dimension and is identifiable. On the other hand, assuming $d_1 \leq d_2 \leq \ldots \leq d_k$, $\mathcal{H}$ is deficient when $\prod_{i=1}^{k-1}d_i - \sum_{i=1}^{k-1}(d_i -1 ) \leq r \leq \min\left\{ d_k, \prod_{i=1}^{k-1}d_i - 1\right\}$. Finally, under the same conditions, $\mathcal{H}$ is identifiable when $\frac{1}{2} \sum_{i} (d_i - 1) + 1 \geq \max \{ d_k, r \}$. Obtaining bounds and results of this type is highly non-trivial and is an open area of research.

In the remainder of the paper, we will focus on simpler latent class models for tables of dimension $k=2$ and illustrate with examples the results mentioned above. For latent class models on two-way tables, there is an alternative,  quite convenient way of describing $\mathcal{H}$ by representing each ${\bf p}$ in $\Delta_{d-1}$ as a $d_1 \times d_2$ matrix and by interpreting the map $\sigma$ as a vector product. In fact, each point ${\bf p}$ in $\mathcal{S}$ is a rank one matrix obtained as  ${\bf p}_1 {\bf p}_2^{\top} $, where ${\bf p}_1 \in \Delta_{d_1 -1}$ and ${\bf p}_2 \in \Delta_{d_1 -2}$ are the appropriate marginal distributions of $X_1$ and $X_2$. Then, the accounting equations for a latent class models with $r$-level become
\[
{\bf p} = \sum_h {\bf p}^{(h)}_1 ({\bf p}^{(h)}_2)^{\top} \lambda_h, \quad \quad \left({\bf p}_1 , {\bf p}_2 , (\lambda_1, \ldots, \lambda_r) \right)\in \Delta_{d_1 -1} \times \Delta_{d_2 -1} \times \Delta_{r-1}
\]
i.e. the matrix ${\bf p}$ is a convex combination of $r$ rank 1 matrices lying on the surface of independence. Therefore all points in $\mathcal{H}$ are non-negative matrices  with entries summing to one and with rank at most $r$. This simple observation allows one to compute the effective dimension of $\mathcal{H}$ for 2-way table as follows. In general, a real valued $d_1 \times d_2$ matrix has rank $r$ or less if and only if the homogeneous polynomial equations corresponding to all of its $(r+1) \times (r+1)$ minors all vanish. Provided $k < \min\{ d_1,d_2 \}$, on $\mathbb{R}^{d_1} \times \mathbb{R}^{d_2}$, the zero locus of all such such equations form a \textit{determinantal} variety of co-dimension $(d_1 - r) (d_2 - r)$ \citep[Proposition 12.2]{HARRIS:92} and hence has dimension $r(d_1 + d_2) - r^2$. Subtracting this value from the expected dimension computed above, and taking into account the fact that all the points lie inside the simplex, we obtain
\begin{equation}\label{eq:2way}
r (d_1 + d_2 -2) + r - 1 - \left( r(d_1 + d_2) - r^2 -1 \right) = r(r-1).
\end{equation}
This number is also the difference between the dimension of the (fully identifiable, i.e. of expected dimension) graphical model of conditional independence $X_1$ and $X_2$ given $H$, 
and the deficient dimension of the latent class model obtained by marginalizing over the variable $H$. 

The study of higher dimensional tables is still an open area of research. The mathematical machinery required to handle larger dimensions is considerably more complicated and relies on the notions higher-dimensional tensors, rank tensors and non-negative rank tensors, for which only partial results exist. See \cite{KRUSKAL:75}, \cite{CHROH:93} and \cite{STARSSEN:83} for details. Alternatively, \cite{MSVS:03} conduct an algebraic-topological investigation of the topological  properties of stochastic factorization of stochastic matrices representing models of conditional independence with one hidden variable and \cite{AR:06,AR:07} explore an overlapping set of problems framed in the context of trees with latent nodes and branches.

The specific case of $k$-way tables with $2$ level latent variables is a fortunate exception, for which the results for 2-way tables just described apply. In fact, \cite{LM:04} show that that these models are the same as the corresponding model for any two-dimensional table obtained by any ``flattening" of the $d_1 \times \ldots \times d_k$-dimensional array of probabilities ${\bf p}$ into a two-dimensional matrix. Flattening simply means collapsing the $k$ variables into two new variables with $f_1$ and $f_2$ levels, and re-organizing the entries of the $k$-dimensional tensor ${\bf p} \in \Delta_{d-1}$ into a $f_1 \times f_1$ matrix accordingly, where, necessarily,  $f_1 + f_2 = \sum_i d_i$. 
Then,  $\mathcal{H}$ is the determinantal variety which is the zero set of all $3 \times 3$ sub-determinants of the matrix obtained by any such flattening. The second example in Section \ref{sec:symbcomp} below illustrates this result.

\section{Examples Involving Synthetic Data}
We further elucidate the non-identifiability phenomenon from the algebraic and geometric point of view, and the multi-modality of the log-likelihood function issue using few, small synthetic examples. In particular, in the ``100 Swiss Frank" problem below, we embark on a exhaustive study of a table with symmetric data and describe the effects of such symmetries on both the parameter space and the log-likelihood function. Although this example involves one of the simplest cases of LC models, these tables already exhibit considerable statistical and geometric complexity.

\subsection{Effective Dimension and Polynomials}\label{sec:symbcomp}
 


We show how it is possible to take advantage of the polynomial nature of equations (\ref{eq:margmap}) to gain further insights into the algebraic properties of distributions obeying latent class models. 
All the computations that follow were made in {\tt SINGULAR} \citep{SINGULAR:05} and are described in details, along with more examples, in \cite{YI:07}.  Although in principle symbolic algebraic software allows one to compute the set of polynomial equations that fully characterize LC models and their properties, this is still a rather difficult and costly task that can be accomplished only for smaller models.

The accounting equations (\ref{eq:margmap}) determine a polynomial mapping $f \colon \Theta \rightarrow \Delta^{d-1}$ given by
\begin{equation}\label{eq:polymap}
(p_1(i_1) \ldots  p_k(i_k), \lambda_h)  \mapsto   \sum_{h \in [r]} p_1(i_1) \ldots  p_k(i_k) \lambda_h,
\end{equation}
so that the latent class model can be analytically defined as the image of this map, i.e. $\mathcal{H} = f(\Theta)$. Then, following the geometry-algebra dictionary principle \citep[see, e.g.,][]{CLOS:96}, the problem of computing the effective dimension of $\mathcal{H}$ can in turn be geometrically cast as a problem of computing the dimension of the image of a polynomial map.  We illustrate how this representation offers considerable advantages with some small examples.

Consider a $2 \times 2 \times 2$ table with $r=2$ latent classes.
 From Proposition 2.3 in \cite{CGG:02}, the latent class models with 2 classes and 3 manifest variables are identifiable. 
The standard dimension, i.e. the dimension of the parameter space $\Theta$ is $r \sum_i (d_i - 1) + r - 1  = 7$, which coincides with the dimension of the enveloping simplex $\Delta_7$. Although this condition implies that the number of parameters to estimate is no larger than the number of cells in the table, a case which, if violated, would entail non-identifiability, it does not guarantee that the effective dimension is also $7$. This can be verified by checking that the symbolic rank of the Jacobian matrix of the map (\ref{eq:polymap}) is indeed $7$, almost everywhere with respect to the Lebesgue measure. Alternatively, one can determine the dimension of the non-identifiable subspace using computational symbolic algebra.
First, we define the ideal of polynomials determined by the 8 equations in (\ref{eq:polymap}) in the polynomial ring in which the (redundant) 16 indeterminates are the $8$ joint probabilities in $\Delta_{7}$ and the $3$ pairs of marginal probabilities in $\Delta_1$ for the observable variables, and the marginal probabilities in $\Delta_1$ for the latent variable. Then we use implicization \citep[Chapter 3]{CLOS:96} to eliminate all the marginal probabilities and to study the Groebner basis of the resulting ideal in which the indeterminates are the joint probabilities only. There is only one element in the basis,
\[
p_{111} + p_{112} + p_{121} + p_{122} + p_{211} + p_{212} + p_{221} + p_{222} = 1,
\]
which gives the trivial condition for probability vectors. This implies the map (\ref{eq:polymap}) is surjective, so that  $\mathcal{H} = \Delta_7$ and the effective dimension is also $7$, showing identifiability, at least for positive distributions. 

Next, we consider the $2 \times 2 \times 3$ table with $r=2$. For this model $\Theta$ has dimension $9$ and the image of the mappings  (\ref{eq:polymap}) is $\Delta_9$. The symbolic rank of the associated  Jacobian matrix is $9$ as well and the model is identifiable. The image of the polynomial mapping determined by (\ref{eq:polymap}) is the variety associated to the ideal whose Groebner basis consists of the trivial equaiton
\[
p_{111} + p_{112} + p_{113} + p_{121} + p_{122} + p_{123} + p_{211} + p_{212} + p_{213} + p_{221} + p_{222} + p_{223} = 1,
\]
and four polynomials corresponding to the determinants
\begin{equation}\label{eq:det}
\begin{array}{c}
\left|
\begin{array}{ccc}
	p_{121} & p_{211} & p_{221} \\
	p_{122} & p_{212} & p_{222} \\
	p_{123} & p_{213} & p_{223} \\
\end{array}\right| \\
\;\\
\left|
\begin{array}{ccc}
	p_{1+1} & p_{211} & p_{221} \\
	p_{1+2} & p_{212} & p_{222} \\
	p_{1+3} & p_{213} & p_{223} \\
\end{array}\right| \\
\;\\
\left|
\begin{array}{ccc}
	p_{+11} & p_{121} & p_{221} \\
	p_{+12} & p_{122} & p_{222} \\
	p_{+13} & p_{123} & p_{223} \\
\end{array}\right| \\
\;\\
\left|
\begin{array}{ccc}
	p_{111} & p_{121}+p_{211} & p_{221} \\
	p_{112} & p_{122}+p_{212} & p_{222} \\
	p_{113} & p_{123}+p_{213} & p_{223} \\
\end{array}\right| \\
\end{array}
\end{equation}
where the subscript symbol ``+" indicates summation over that coordinate. 
In turn, the zero set of the above determinants coincide with the determinantal variety specified by the zero set of all $3 \times 3$ minors of the 3$\times$4 matrix
\begin{equation}\label{eq:3x4}
\left(
\begin{array}{cccc}
	p_{111} & p_{121} & p_{211} & p_{221} \\
	p_{112} & p_{122} & p_{212} & p_{222} \\
	p_{113} & p_{123} & p_{213} & p_{223} \\
\end{array}
\right)
\end{equation}
which is a flattening of the $2 \times 2 \times 3$ array of probabilities describing the joint distribution for the latent class model under study. This is in accordance with the result in \cite{LM:04} of mentioned above.  Now, the determinantal variety given by the vanishing locus of all the $3 \times 3$ minors of the matrix (\ref{eq:3x4}) is the latent class model for a $3 \times 4$ table with $2$ latent classes, which, according to (\ref{eq:2way}), has deficiency equal to $2$. The effective dimension of this variety is $9$, computed as the standard dimension, $11$, minus the deficiency. Then,the effective dimension of the model we are interested is also $9$ and we conclude that the model is identifiable.

Table \ref{tab:dim} summarizes some of our numerical evaluations of the different notions of dimension for a different LC models. We computed the effective dimensions  by evaluating with {\tt MATLAB} the numerical rank of the Jacobian matrix, based on the simple algorithm suggested in \cite{GHKM:01} and also using {\tt SINGULAR}, for which  only computations involving small models were feasible.

\begin{table}[ht]
\caption{Different dimensions of some latent class models. The Complete Dimension is the dimension $d - 1$ of the envoloping probability simplex $\Delta_{d-1}$. See also Table 1 in \cite{KZ:02}.}
	\centering
	\begin{tabular}{||c|c||c||c||c||c||}\hline
		\multicolumn{2}{|c||}{ } & Effective  & Standard  & Complete   & \\
		\multicolumn{2}{|c||}{Latent Class Model} &   Dimension &   Dimension &   Dimension & Deficiency  \\\hline
		$\Delta_{d-1}$ & r  &  & & & \\\hline 
		$2\times2$ & $2$  & 3 & 5 & 3 & 0 \\
		$3\times3$ & $2$  & 7 & 9 & 8 & 1\\
		$4\times5$ & $3$  & 17  & 23 & 19 & 2\\
		$2\times2\times2$ & $2$  & 7 & 7 & 7 & 0\\
		$2\times2\times2$ & $3$  & 7 & 11 & 7 & 0\\
		$2\times2\times2$ & $4$  & 7 & 15 & 7 & 0\\
		$3\times3\times3$ & $2$  & 13 & 13 & 26 & 0\\
		$3\times3\times3$ & $3$  & 20 & 20 & 26 & 0\\
		$3\times3\times3$ & $4$  & 25 & 27 & 26 & 1\\
		$3\times3\times3$ & $5$  & 26 & 34 & 26 & 0\\
		$3\times3\times3$ & $6$  & 26 & 41 & 26 & 0\\
		$5\times2\times2$ & $3$  & 17 & 20 & 19 & 2\\
		$4\times2\times2$ & $3$  & 14 & 17 & 15 & 1\\
		$3\times3\times2$ & $5$  & 17 & 29 & 17 & 0\\
		$6\times3\times2$ & $5$  & 34 & 44 & 35 & 1\\
		$10\times3\times2$ & $5$  & 54 & 64 & 59 & 5\\
		$2\times2\times2\times2$ & $2$  & 9 & 9 & 15 & 0\\
		$2\times2\times2\times2$ & $3$  & 13 & 14 & 15 & 1\\
		$2\times2\times2\times2$ & $4$  & 15 & 19 & 15 & 0\\
		$2\times2\times2\times2$ & $5$  & 15 & 24 & 15 & 0\\
		$2\times2\times2\times2$ & $6$  & 15 & 29 & 15 & 0\\
		\hline
	\end{tabular}
	\label{tab:dim}
\end{table}

\subsection{The 100 Swiss Franc Problem}

\subsubsection{Introduction}
Now we study the problem of fitting a non-identifiable 2-level latent class model to a two-way table with symmetry counts. This problem was suggested by Bernd Sturmfels to the participants of his postgraduate lectures on Algebraic Statistics held at ETH Zurich in the Summer semester of 2005 (where he offered 100 Swiss franks for a rigorous solution), and is described in detail as Example 1.16 in \cite{PS:05}. 
The observed table is
\begin{equation}\label{eq:swiss}
n=\left(
\begin{array}{cccc}
4& 2& 2& 2\\
2 &4 &2 &2\\
2& 2 &4 &2\\
2 &2 &2 &4\\
\end{array}\right).
\end{equation}
For the basic latent class  model, the standard dimension of $\Theta = \Delta_3 \times \Delta_3 \times \Delta_1$ is $2(3+3)+1 = 13$ and, by (\ref{eq:2way}), the deficiency is $2$. Thus, the model is not identifiable and the pre-image of each point ${\bf p} \in \mathcal{H}$ by the map (\ref{eq:polymap}) is a 2-dimensional surface in $\Theta$.
To keep the notation light, we write $\alpha_{ih}$ for $p_1^{(h)}(i)$ and $\beta_{jh}$ for $p_2^{(h)}(j)$, where $i,j = 1, \ldots, 4$ and $\alpha^{(h)}$ and $\beta^{(h)}$ for the conditional marginal distribution of $X_1$ and $X_2$ given $H=h$, respectively. The accounting equations for the points in $\mathcal{H}$ become
\begin{equation}\label{eq:lik.ab}
p_{ij} = \sum_{h \in\{1,2\}} \lambda_h \alpha_{ih} \beta_{jh}, \quad \quad i,j \in  [4]
\end{equation}
and the log-likelihood function, ignoring an irrelevant additive constant, is 
\[
\ell(\theta)= \sum_{i,j} n_{ij} \log \left( \sum_{h \in\{1,2\}} \lambda_h \alpha_{ih} \beta_{jh} \right), \quad \quad \theta \in \Delta_{3} \times \Delta_3 \times \Delta_1.
\]
It is worth emphasizing, as we did above and as the previous display clearly shows, that the observed counts are minimal sufficient statistics.

Alternatively, we can re-parametrize the log-likelihood function using directly points in $\mathcal{H}$ rather the points in the parameter space $\Theta$. Recall from our discussion in section \ref{sec:geom} that, for this model, the $4 \times 4$ array ${\bf p}$ is in $\mathcal{H}$ if and only if each $3 \times 3$ minor vanishes. Then, we can write the log-likelihood function as
\begin{equation}\label{eq:lik.p}
\ell({\bf p}) = \sum_{i,j} n_{ij} \log p_{ij}, \quad \quad {\bf p} \in \Delta_{15}, \,\,\, \m{det}({\bf p}^*_{ij}) = 0  \; \m{all} \;  i,j \in [4],
\end{equation}
where ${\bf p}^*_{ij}$ is the $3 \times 3$ sub-matrix of ${\bf p}$ obtained by erasing the $i$th row and the $j$th column.

Although the first order optimality conditions for the Lagrangian corresponding to the parametrization  (\ref{eq:lik.p}) are algebraically simpler and can be given the form of a system of a polynomial equations, in practice, the classical parametrization (\ref{eq:lik.ab}) is used in both the EM and the Newton-Raphson implementations in order to compute the maximum likelihood estimate of ${\bf p}$. See \cite{GOODMAN:79}, \cite{HAB:88}, and \cite{RW:84} for more details about these numerical procedures.


\subsubsection{Global and Local Maxima}
Using both EM and Newton-Raphson algorithm with several different starting points, we found $7$ local maxima of the log-likelihood function, reported in Table \ref{tab:maxima}. The global maximum was found experimentally to be $-20.8074 + const.$, where $const.$ denotes the additive constant stemming from the multinomial coefficient.  The maximum is achieved by the three tables of fitted values Table \ref{tab:maxima} {\bf a)}. The remaining four tables  are  local maximum of $-20.8616 + const.$, close in value to the actual global maximum.
Using {\tt SINGULAR} (see~\citep{SINGULAR:05}), we checked that the tables found satisfy the first order optimality conditions (\ref{eq:lik.p}). After verifying numerically the second order optimality conditions, we conclude that those points are indeed local maxima. Furthermore, as indicated in \cite{PS:05}, the log-likelihood function also has a few saddle points. 

A striking feature of the global maxima in Table \ref{tab:maxima} is their invariance under the action of the symmetric group on four elements acting simultaneously on the row and columns. Different symmetries arise for the local maxima. We will give an explicit representation of these symmetries under the classical parametrization (\ref{eq:lik.ab}) in the  next section.

Despite the simplicity and low-dimensionality of the LC model for the Swiss franc problem and the strong symmetric features of the data, we have yet to provide a purely mathematical proof that the three top arrays in Table \ref{tab:maxima} correspond to a global maximum of the likelihood function. We view the difficulty and complexity of the 100 Swiss Francs problem as a consequence of the inherent difficulty of even small LC models and perhaps an indication that the current theory has still many open, unanswered problems.  In Section \ref{sec:proof}, we present  partial results towards the completion of the proof.


\begin{table}
\caption{Tables of fitted value corresponding to $7$ the maxima of the likelihood equation for the observed table (\ref{eq:swiss}). {\bf a)}: global maximua (log-likelihood value $-20.8079$). {\bf b)}: local maxima (log-likelihood value $-20.8616$).}
\[
\begin{array}{c}
{\bf a)}\\
\end{array}
\]
\[
\begin{array}{ccc}
\left(\begin{array}{cccc}
              3 &3 &2 &2\\
              3 &3 &2 &2\\
              2 &2 &3 &3\\
              2 &2 &3 &3\\
\end{array}\right) &
\left(
\begin{array}{cccc}
              3 &2 &3 &2\\
              2 &3 &2 &3\\
              3 &2 &3 &2\\
              2 &3 &2 &3\\
\end{array}\right) &
\left(
\begin{array}{cccc}
              3 &2 &2 &3\\
              2 &3 &3 &2\\
              2 &3 &3 &2\\
              3 &2 &2 &3\\
\end{array}\right)
\end{array}
\]
\[
\begin{array}{c}
{\bf b)}\\
\end{array}
\]
\[
\begin{array}{cc}
\left(\begin{array}{cccc}
              8/3 &8/3 &8/3 &2\\
              8/3 &8/3 &8/3 &2\\
              8/3 &8/3 &8/3 &2\\
              2 &2 &2 &4\\
\end{array}\right) &
\left(
\begin{array}{cccc}
              8/3 &8/3 &2 &8/3\\
              8/3 &8/3 &2 &8/3\\
              2 &2 &4 &2\\
              8/3 &8/3 &2 &8/3\\
\end{array}\right) \\
\left(
\begin{array}{cccc}
              8/3 &2 &8/3 &8/3\\
              2 &4 &2 &2\\
              8/3 &2 &8/3 &8/3\\
              8/3 &2 &8/3 &8/3\\
\end{array}\right) &
\left(
\begin{array}{cccc}
              4 &2 &2 &2\\
              2 &8/3 &8/3 &8/3\\
              2 &8/3 &8/3 &8/3\\
              2 &8/3 &8/3 &8/3\\
\end{array}\right)
\end{array}
\]
\label{tab:maxima}
\end{table}

\subsubsection{Unidentifiable Space}
It follows from equation (\ref{eq:2way}) that the non-identifiable subspace is a two-dimensional subset of $\Theta$. We give  an explicit algebraic description of this space, which we will then use to obtain interpretable plots of the profile likelihood.  

  \begin{figure}[t]
\centering
\includegraphics[width=3in]{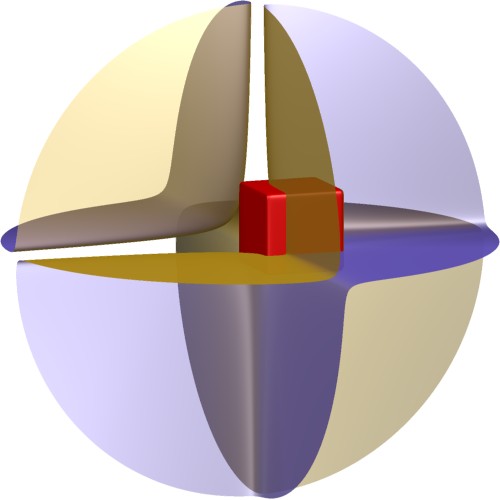}
\caption{The 2-dimensional surface defined by equation (\ref{eq:fixedsurf}), when evaluated over the ball in $\mathbb{R}^3$ of radius $3$, centered at the origin. The inner box is the unit cube $[0,1]^3$.}
\label{fig:mlesurface}
\end{figure}

\begin{figure}[t]
\centering
\begin{tabular}{c}
{\bf a)}\\
 \includegraphics[width=3in]{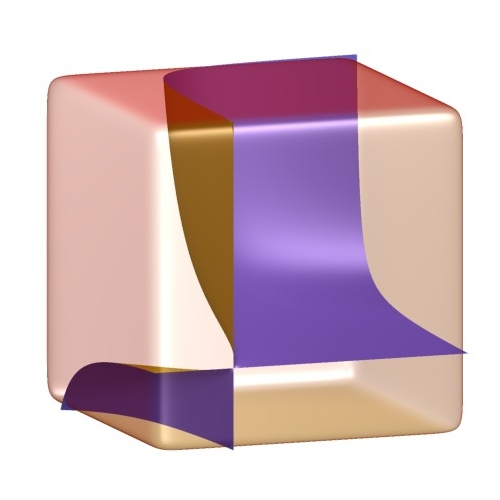} \\
 {\bf b)}\\
 \includegraphics[width=3.5in]{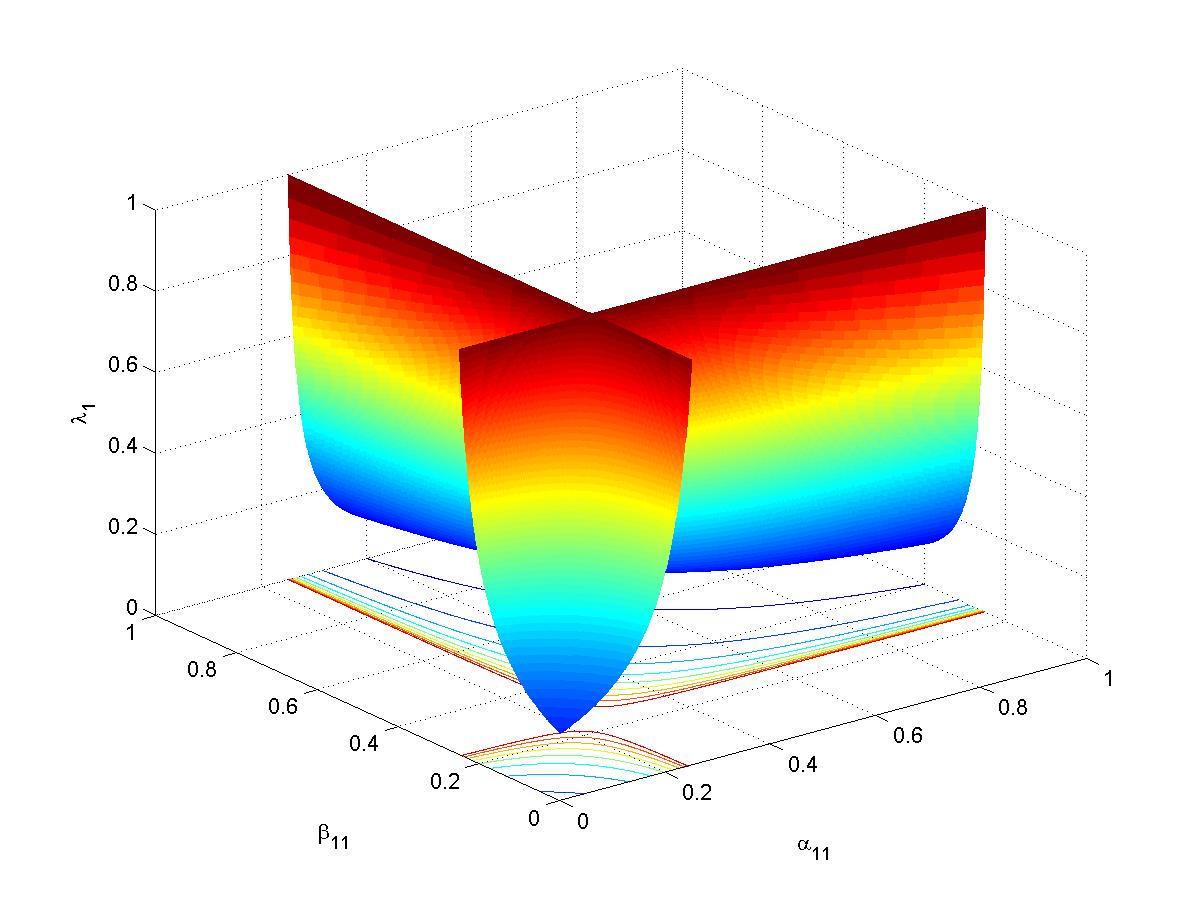} \\
\end{tabular}
\caption{Intersection of the surface defined by equation (\ref{eq:fixedsurf}) with the unit cube $[0,1]^3$, different views obtained using {\tt surf} in {\bf a)}  and {\tt MATLAB} in  {\bf b)}.}
\label{fig:idealcube}
\end{figure}

Firstly, we focus on the three global maxima in Table \ref{tab:maxima} {\bf a)}. 
By the well-known properties of the EM algorithm \citep[see, e.g.,][Theorem 1.15]{PS:05}, if the vector of parameters $\theta$  is a stationary point in the maximization step of the EM algorithm, then $\theta$ is a critical point and hence a good candidate for a local maximum. Using this observation, it is possible to show \citep[see][]{YI:07}  that any point in $\Theta$ satisfying the equations 
\begin{equation}
\begin{array}{l}
	\alpha_{1h}=\alpha_{2h}, \; \alpha_{3h}=\alpha_{4h} \quad h=1,2\\
	\beta_{1h}=\beta_{2h},  \;\beta_{3h}=\beta_{4h} \quad h=1,2 \\
	\sum_h\lambda_h\alpha_{1h}\beta_{1h} = \sum_h\lambda_h\alpha_{3h}\beta_{3t} = 3/40\\
	\sum_h\lambda_h\alpha_{1h}\beta_{3h} = \sum_h\lambda_h\alpha_{3h}\beta_{1t} = 2/40\\
\end{array}	
\label{eq:fixedpt}
\end{equation}
is a stationary point. Notice that the first four equations in (\ref{eq:fixedpt}) require $\alpha^{(h)}$ and $\beta^{(h)}$ to each have the first and second pairs of coordinates identical, for $h=1,2$.  
The equation (\ref{eq:fixedpt}) defines a 2-dimensional surface in $\Theta$. Using {\tt SINGULAR}, we can verify that, holding, for example, $\alpha_{11}$ and $\beta_{11}$ fixed, determines all of  the other parameters according to the equations
\[
\left\{
\begin{array}{l}
\lambda_1=\frac{1}{80\alpha_{11}\beta_{11}-20\alpha_{11}-20*\beta_{11}+6} \\
\lambda_2=1-\lambda_1 \\
\alpha_{21}=\alpha_{11} \\
\alpha_{31}=\alpha_{41} = 0.5 - \alpha_{11}\\
\alpha_{12} = \alpha_{22}=\frac{10\beta_{11}-3}{10(4\beta_{11}-1)} \\\
\alpha_{32} = \alpha_{42} = 0.5-\alpha_{12} \\
\beta_{21}=\beta_{11}\\
\beta_{31}=\beta_{41}=0.5-\beta_{11}\\
\beta_{12}=\beta_{22}=\frac{10\alpha_{11}-3}{10(4\alpha_{11}-1)} \\
\beta_{32}=\beta_{42}=0.5-\beta_{12}.\\
\end{array}\right.
\]
Using \textit{elimination} \citep[see][Chapter 3]{CLOS:96} to remove all the variables in the system except for $\lambda_1$, we are left with one equation
\begin{equation}\label{eq:fixedsurf}
80 \lambda_1 \alpha_{11} \beta_{11} - 20 \lambda_1 \alpha_{11} - 20 \lambda_1 \beta_{11} + 6 \lambda_1 - 1 =0.
\end{equation}
Without the constraints for the coordinates of $\alpha_{11}$, $\beta_{11}$ and $\lambda_1$ to be probabilities,  (\ref{eq:fixedsurf}) defines a two-dimensional object in $\mathbb{R}^3$, depicted in Figure \ref{fig:mlesurface}. Notice that the axes do not intersect this surface, so that zero is not a possible value for $\alpha_{11}$, $\beta_{11}$ and $\lambda_1$. 
Because the non-identifiable space in $\Theta$ is 2-dimensional, equation (\ref{eq:fixedsurf}) actually defines a bijection between $\alpha_{11}$, $\beta_{11}$ and $\lambda_1$ and the rest of the parameters. Then, the intersection of the surface (\ref{eq:fixedsurf}) with the unit cube $[0,1]^3$, depicted as a red box in Figure \ref{fig:mlesurface}, is the projection of the whole non-identifiable subspace into the 3-dimensional unit cube where $\alpha_{11}$, $\beta_{11}$ and $\lambda_1$ live.
Figure \ref{fig:idealcube} displays two different views of this projection.


\begin{figure}[h]
\centering
\includegraphics[width=3in]{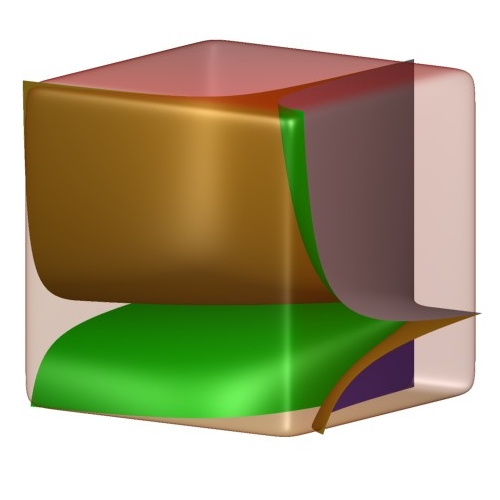}
\caption{Projection of the non-identifiable spaces corresponding to the first and second and third MLE from Table \ref{tab:maxima} {\bf a)} into the 3-dimensional unit cube where $\lambda_1$, $\alpha_{11}$ and $\beta_{21}$ take values.}
\label{fig:idealcube2}
\end{figure}
 
 \begin{figure}[h]
\centering
\includegraphics[width=3in]{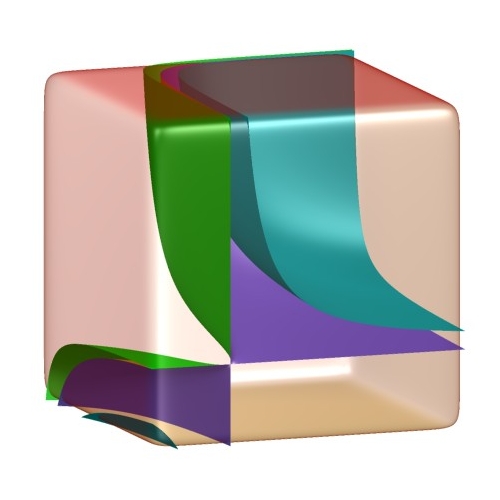}
\caption{Projection of the non-identifiable spaces the first MLE in Table \ref{tab:maxima} {\bf a)}, the first three local maxima and the last local maxima in Table \ref{tab:maxima} {\bf b)} into the 3-dimensional unit cube where $\lambda_1$, $\alpha_{11}$ and $\beta_{11}$ take values. In this coordinate system, the projection of non-identifiable subspaces for the first three local maxima in Table \ref{tab:maxima} {\bf b)} results in the same surface; in order to obtain distinct surfaces, it would be necessary to change the coordinates over which the projections are made.}
\label{fig:idealcube3}
\end{figure}

The preceding arguments hold unchanged if we replace the symmetry conditions in the first two lines of equation (\ref{eq:fixedpt}) with either of these other two conditions, requiring different pairs of coordinates to be identical, namely
\begin{equation}\label{eq:fixedpt2}
	\alpha_{1h}=\alpha_{3h}, \; \alpha_{2h}=\alpha_{4h},\;\;
	\beta_{1h}=\beta_{3h},  \;\beta_{2h}=\beta_{4h}\\
\end{equation}
and
\begin{equation}\label{eq:fixedpt3}
	\alpha_{1h}=\alpha_{4h}, \; \alpha_{2h}=\alpha_{3h},\;\;
	\beta_{1h}=\beta_{4h},  \;\beta_{2h}=\beta_{3h}, \\
\end{equation}
where $h=1,2$.

By our computations, the non-identifiable surfaces inside $\Theta$ corresponding each to one of the three pairs of coordinates held fixed in equations (\ref{eq:fixedpt}), (\ref{eq:fixedpt2}) and (\ref{eq:fixedpt3}), produce the three distinct tables of maximum likelihood estimates reported in Table \ref{tab:maxima} {\bf a)}.  Figure \ref{fig:idealcube} shows the projection of the non-identifiable subspaces for the three MLEs in Table \ref{tab:maxima} {\bf a)} into the three dimensional unit cube for $\lambda_1$, $\alpha_{11}$ and $\beta_{11}$. Although each of these three subspaces are disjoint subsets of $\Theta$, their lower dimensional projections comes out as unique. By projecting onto the different coordinates $\lambda_1$, $\alpha_{11}$ and $\beta_{21}$ instead, we obtain two disjoint surfaces for the first, and second and third MLE, shown in Figure \ref{fig:idealcube2}.

Table \ref{tab:estpar} presents some estimated parameters using the EM algorithm. Though these estimates are hardly meaningful, because of the non-identifiability issue, they show the symmetry properties we pointed out above and implicit in equations  (\ref{eq:fixedpt}), (\ref{eq:fixedpt2}) and (\ref{eq:fixedpt3}), and they explain the invariance under simultaneous permutation of the fitted tables. In fact, the number of global maxima is the number of different configurations of the 4 dimensional vectors of estimated marginal probabilities with two identical coordinates, namely $3$. This phenomenon, entirely due to the strong symmetry in the observed table (\ref{eq:swiss}),  is completely separate from the non-identrifiability issues, but just as  problematic.

By the same token, we can show that vectors of marginal probabilities with $3$  identical coordinates also produce stationary points for the EM algorithms. This type of stationary points trace surfaces inside $\Theta$ which determine the local maxima of Table \ref{tab:maxima} {\bf b)}. The number of these local maxima corresponds, in fact, to the number of possible configurations of 4-dimensional vectors with $3$ identical  coordinates, namely $4$. Figure \ref{fig:idealcube3} depicts the lower dimensional projections into $\lambda_1$, $\alpha_{11}$ and $\beta_{11}$ of the non-identifiable subspaces for the first MLE in Table \ref{tab:maxima} {\bf a)}, the first three local maxima and the last local maxima in Table \ref{tab:maxima} {\bf b)}.  

We can summarize our finding as follows: the maxima in Table \ref{tab:maxima} define  disjoint 2-dimensional surfaces inside the parameter space $\Theta$, the projection of one of them depicted in Figure \ref{fig:idealcube}. While non-identifiability is a structural feature of these models which is independent of the observed data, the multiplicity and invariance properties of the maximum likelihood estimates and the other local maxima is a phenomenon cause by the symmetry in the observed table of counts.

\begin{table}
\caption{Estimated parameters by the EM algorithm for the three global maxima in Table \ref{tab:maxima} {\bf a)}.}
\begin{tabular}{||c||ccc||}
\hline
{\bf Estimated Means} & \multicolumn{3}{c||}{\bf  Estimated Parameters }\\
\hline
$ 
\left( \begin{array}{cccc}
              3 &3 &2 &2\\
              3 &3 &2 &2\\
              2 &2 &3 &3\\
              2 &2 &3 &3\\
\end{array} \right) 
$ & 

$
\widehat{\alpha}^{(1)} = \widehat{\beta}^{(1)} = \left( \begin{array}{c}
0.3474\\
     0.3474\\
     0.1526\\
     0.1526\\
\end{array} \right)
$ & 

$
\widehat{\alpha}^{(2)} = \widehat{\beta}^{(2)} = \left( \begin{array}{c}
0.1217\\
     0.1217\\
     0.3783\\
     0.3783\\
\end{array} \right)
$ &

$
\widehat{\lambda} =\left(\begin{array}{c} 0.5683\\ 0.4317\end{array}\right)
$\\
$\;$ & $\;$  & $\;$ & $\;$  \\
$
\left(
\begin{array}{cccc}
              3 &2 &3 &2\\
              2 &3 &2 &3\\
              3 &2 &3 &2\\
              2 &3 &2 &3\\
\end{array}\right) 
$ & 
$
\widehat{\alpha}^{(1)} = \widehat{\beta}^{(1)} = \left( \begin{array}{c}
0.3474\\
0.1526\\
0.3474\\
0.1526\\
\end{array} \right)
$ &

$  
\widehat{\alpha}^{(2)} = \widehat{\beta}^{(2)} = \left( \begin{array}{c}
0.1217\\
0.3783\\
0.1217\\
0.3783\\
\end{array} \right)
$ &

$
\widehat{\lambda} =\left(\begin{array}{c} 0.5683\\ 0.4317\end{array}\right)
$\\
$\;$ & $\;$  & $\;$ & $\;$  \\
$
\left(
\begin{array}{cccc}
              3 &2 &2 &3\\
              2 &3 &3 &2\\
              2 &3 &3 &2\\
              3 &2 &2 &3\\
\end{array}\right) 
$ & 

$
\widehat{\alpha}^{(1)} = \widehat{\beta}^{(1)} = \left( \begin{array}{c}
0.3474\\
0.1526\\
0.1526\\
0.3474\\
\end{array} \right)
$ &

$  
\widehat{\alpha}^{(2)} = \widehat{\beta}^{(2)} = \left( \begin{array}{c}
0.1217\\
0.3783\\
0.3783\\
0.1217\\
\end{array} \right)
$ &

$
\widehat{\lambda} =\left(\begin{array}{c} 0.5683\\ 0.4317\end{array}\right)
$\\

\hline
\end{tabular}
\label{tab:estpar}
\end{table}

\subsubsection{Plotting the Log-likelihood Function}

\begin{figure}[ht]
\centering
\includegraphics[width=4in]{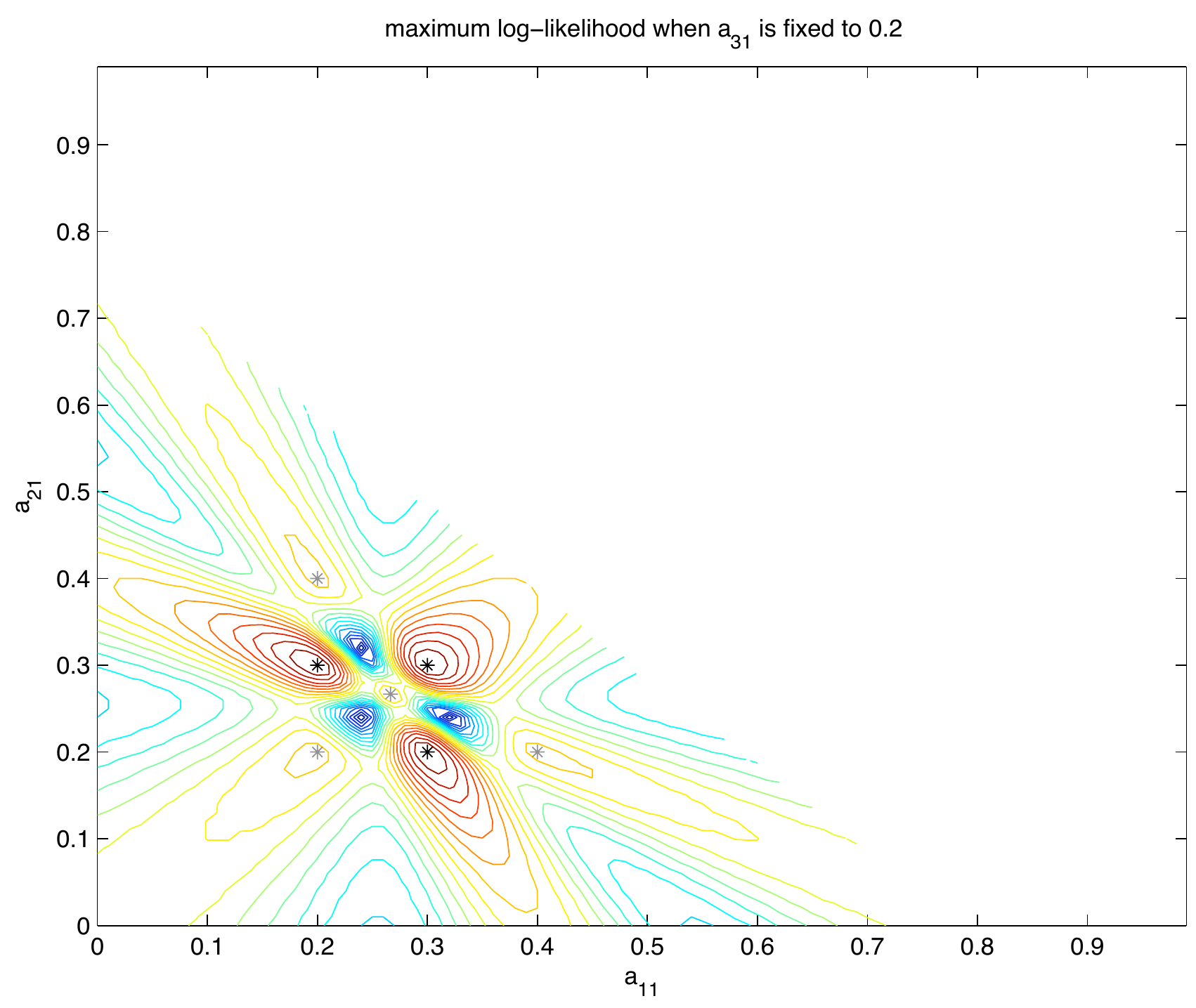}
\caption{The contour plot of the profile likelihood as a function of  $\alpha_{11}$ and $\alpha_{21}$ when $\alpha_{31}$ is fixed. There are seven peaks: the three black points are the MLEs and the four gray points are the other local maxima.}
\label{fig:contour}
\end{figure}

Having determined that the non-identifiable space is 2-dimensional and that there are multiple maxima, we proceed with some plots of the  profile log-likelihood function. To obtain a non-trivial surface, we need to consider three parameters. Figures \ref{fig:mesh} and \ref{fig:contour} display the surface and contour plot of the profile log-likelihhod function for $\alpha_{11}$ and $\alpha_{21}$ when $\alpha_{31}$ is one of the fixed parameters. Both Figures show clearly the different maxima of the log-likelihood function, each lying on the top of ``ridges" of the log-likelihood surface which are placed symmetrically with respect to each others. The position and shapes of these ridges reflect, once again, the invariance properties of the estimated probabilities and parameters.

\begin{figure}[ht]
\centering
\includegraphics[width=5in]{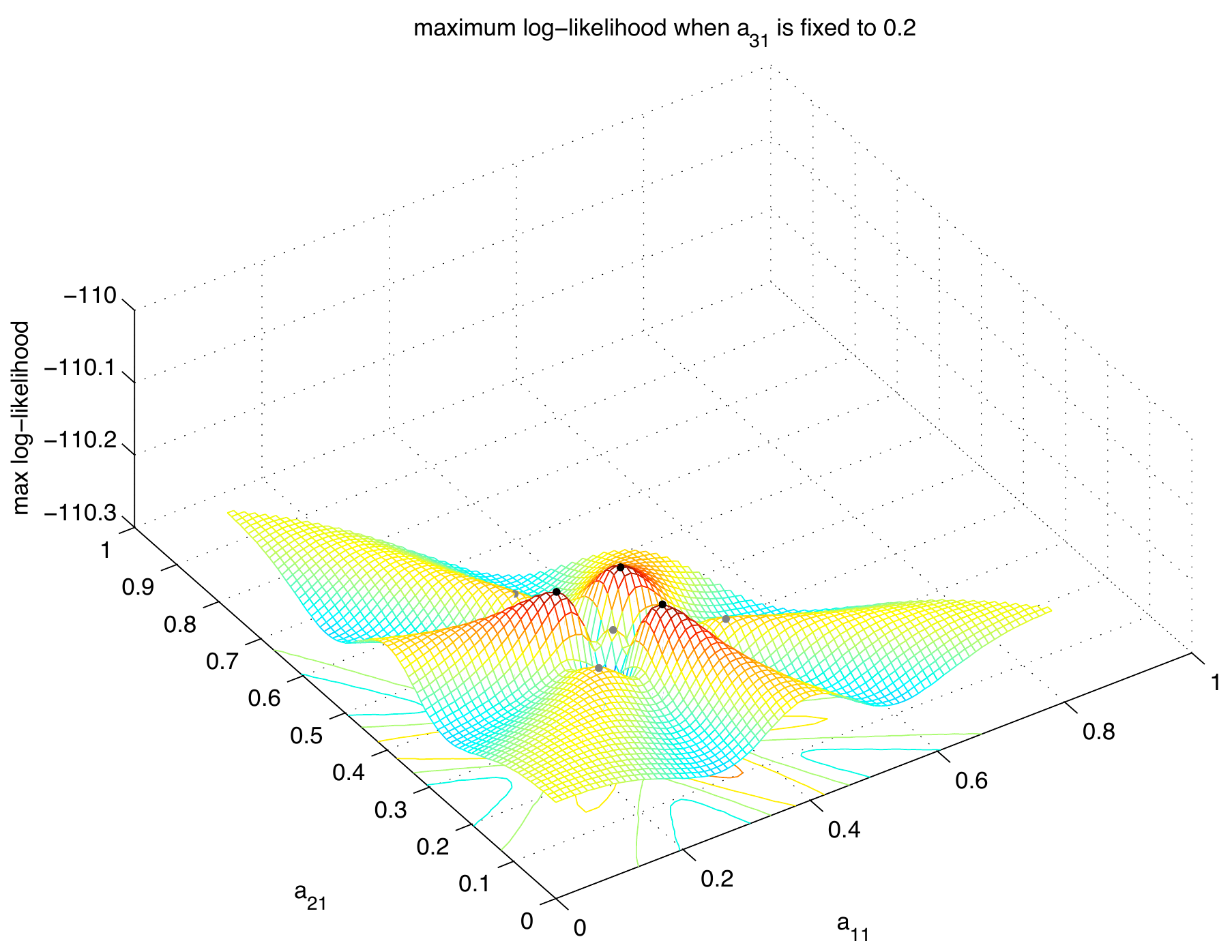}
\caption{The plot of the profile likelihood as a function of $\alpha_{11}$ and $\alpha_{21}$ when $\alpha_{31}$ is fixed to $0.2$. There are seven peaks: the three black points are the MLEs and the four gray diamonds are the other local maxima.}
\label{fig:mesh}
\end{figure}

\subsubsection{Further Remarks and Open Problem}

We conclude this section with some observations and pointers to open problems. 

\begin{figure}
\centering
\includegraphics[width=4in]{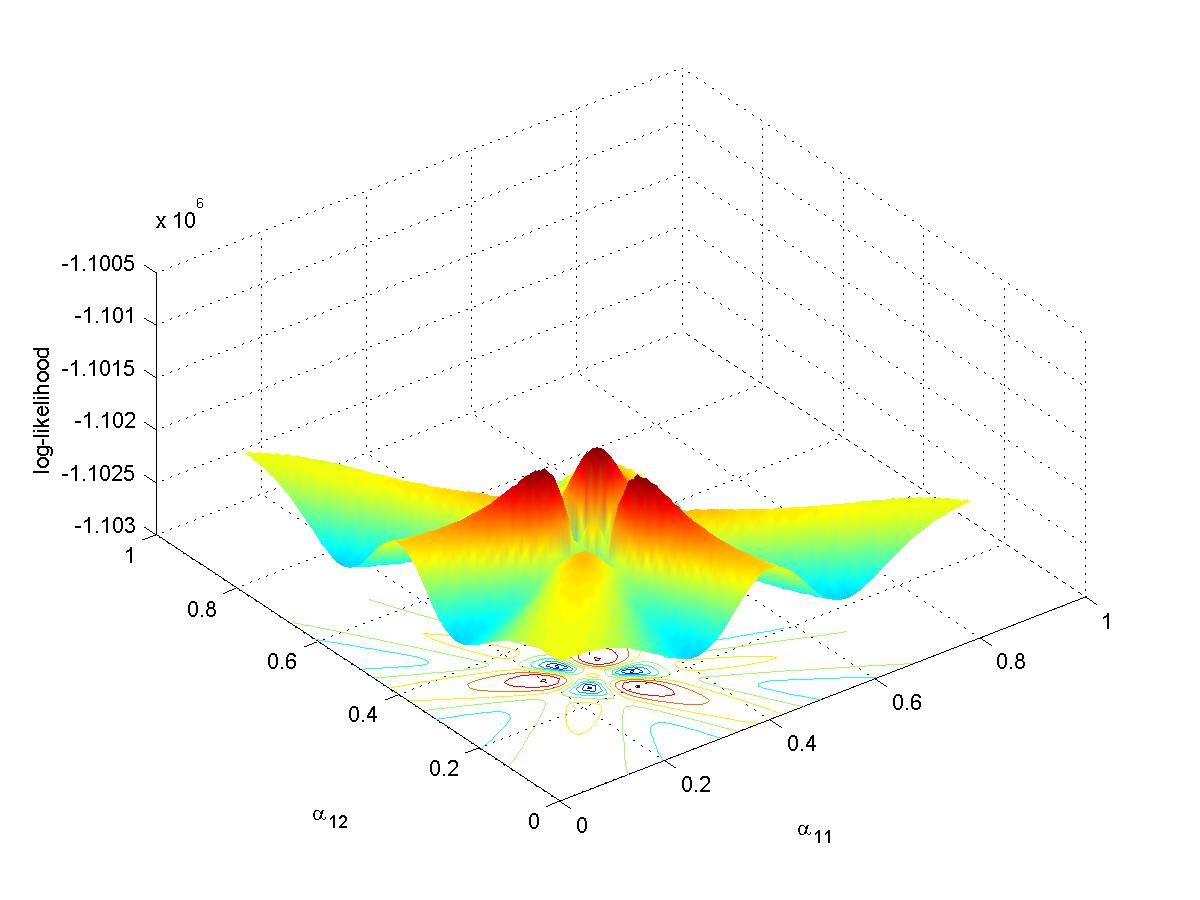}
\caption{The contour plot of the profile likelihood as a function of $\alpha_{11}$ and $\alpha_{21}$ when $\alpha_{31}$ is fixed for the data (\ref{eq:swiss}) multiplied by 10000.
As before, there are seven peaks: three global maxima and four identical local maxima.}
\label{fig:contour10000}
\end{figure}

One of the interesting aspects we came across while fitting the table (\ref{eq:swiss}) was the proximity of the values of the local and global maxima of the log-likelihood function. Furthermore, although these values are very close, the fitted tables corresponding to global and local maxima are remarkably different. Even though the data (\ref{eq:swiss}) are not sparse, we wonder about the effect of cell sizes.
Figure \ref{fig:contour10000} show the same profile log-likelihood for the table (\ref{eq:swiss}) multiplied by 10000. While the number of global and local maxima, the contour plot and the basic symmetric shape of the profile log-likelihood surface remain unchanged after this rescaling, the peaks around the global maxima have become much more pronounced and so has the difference between of the values of the global and local maxima. 

We have studied at a number of variations of table  (\ref{eq:swiss}), focussing in particular on the symmetric data. We report only some of our results and refer to \cite{YI:07} for a more extensive study. Table \ref{tab:mle6x6} shows the values and number of local and global maxima for a the $6 \times 6$ version of (\ref{eq:swiss}). As for the $4 \times 4$ case, we notice strong invariance features of the various maxima of the likelihood function and a very small difference between the value of the global and local maxima. 

\begin{table}[!ht]
\caption{Stationary points for the 6$\times$6 version of the table (\ref{eq:swiss}). All the maxima are invariant under simultaneous permutations of the rows and columns of the corresponding fitted tables.}
\centering
\begin{tabular}{||c|c||}\hline
Fitted counts & Log-likelihood\\
\hline 
 & \\
$
\left(
\begin{array}{cccccc}
4 & 2 & 2 & 2 & 2 & 2\\
2 & 12/5 & 12/5 & 12/5 & 12/5 & 12/5\\
2 & 12/5 & 12/5 & 12/5 & 12/5 & 12/5\\
2 & 12/5 & 12/5 & 12/5 & 12/5 & 12/5\\
2 & 12/5 & 12/5 & 12/5 & 12/5 & 12/5\\
2 & 12/5 & 12/5 & 12/5 & 12/5 & 12/5\\
\end{array}
\right)
$
& $-300.2524 + const.$ \\
 & \\
$
\left(
\begin{array}{cccccc}
   7/3&   7/3&   7/3&   7/3&   7/3&   7/3\\
   7/3&  13/5&  13/5&  13/5& 29/15& 29/15\\
   7/3&  13/5&  13/5&  13/5& 29/15& 29/15\\
   7/3&  13/5&  13/5&  13/5& 29/15& 29/15\\
   7/3& 29/15& 29/15& 29/15& 44/15& 44/15\\
   7/3& 29/15& 29/15& 29/15& 44/15& 44/15\\
\end{array}
\right)
$
& $ -300.1856 + const.$ \\
 & \\
$
\left(
\begin{array}{cccccc}
   3&   3&   2&   2&   2&   2\\
   3&   3&   2&   2&   2&   2\\
   2&   2& 5/2& 5/2& 5/2& 5/2\\
   2&   2& 5/2& 5/2& 5/2& 5/2\\
   2&   2& 5/2& 5/2& 5/2& 5/2\\
   2&   2& 5/2& 5/2& 5/2& 5/2\\
\end{array}
\right)
$
& $-300.1729 + const.$ \\
 & \\
$
\left(
\begin{array}{cccccc}
 8/3& 8/3& 8/3&   2&   2&   2\\
 8/3& 8/3& 8/3&   2&   2&   2\\
 8/3& 8/3& 8/3&   2&   2&   2\\
   2&   2&   2& 8/3& 8/3& 8/3\\
   2&   2&   2& 8/3& 8/3& 8/3\\   
   2&   2&   2& 8/3& 8/3& 8/3\\
\end{array}
\right)
$
&$-300.1555 + const.$ (MLE) \\
 & \\
$
\left(
\begin{array}{cccccc}
 7/3& 7/3& 7/3&  7/3&   7/3&   7/3\\
 7/3& 7/3& 7/3&  7/3&   7/3&   7/3\\
 7/3& 7/3& 7/3&  7/3&   7/3&   7/3\\
 7/3& 7/3& 7/3&  7/3&   7/3&   7/3\\
 7/3& 7/3& 7/3&  7/3&   7/3&   7/3\\
 7/3& 7/3& 7/3&  7/3&   7/3&   7/3\\\end{array}
\right)
$
& $-301.0156 + const.$ \\
 & \\
$
\left(
\begin{array}{cccccc}
7/3 & 7/3 & 7/3 & 7/3 & 7/3 & 7/3 \\
7/3 & 35/9 & 35/18 & 35/18 & 35/18 & 35/18 \\
7/3 & 35/18 & 175/72 & 175/72 & 175/72 & 175/72 \\
7/3 & 35/18 & 175/72 & 175/72 & 175/72 & 175/72 \\
7/3 & 35/18 & 175/72 & 175/72 & 175/72 & 175/72 \\
7/3 & 35/18 & 175/72 & 175/72 & 175/72 & 175/72 \\
\end{array}
\right)
$
& $-300.2554 + const.$ \\
 & \\
\hline
\end{tabular}
\label{tab:mle6x6}
\end{table}

Fitting the same model to the table
\[
\left(
\begin{array}{cccc}
1 & 2 & 2 & 2\\
2 & 1 & 2 & 2\\
2 & 2 & 1 & 2\\
2 & 2 & 2 & 1\\
\end{array}
\right)
\]
we found 6 global maxima of the likelihood function, which give as many maximum likelihood estimates, all obtainable via simultaneous permutation of rows and columns of the table
\[
\begin{array}{ll}
\left(
\begin{array}{cccc}
7/4 & 7/4 & 7/4 & 7/4 \\
7/4 & 7/4 & 7/4 & 7/4 \\
7/4 & 7/4 & 7/6 & 7/3 \\
7/4 & 7/4 & 7/3 & 7/6 \\
\end{array}
\right), &
\textnormal{log-likelihood} = -77.2927 + const.
\end{array}
\]

Based on the various cases we have investigated, we have the following conjecture, which we verified computationally up to dimension $k = 50$:

\paragraph{Conjecture:}
The MLEs 
For the $n\times n$ table with values $x$ along the diagonal and values $y \leq x$ for off the diagonal elements, the maximum likelihood estimates for the latent class model with $2$ latent classes are the 2$\times$2 block diagonal matrix of the form 
$\left(\begin{array}{cc}A & B \\ B' & C\end{array}\right)$
and the permutated versions of it,
where $A$, $B$, and $C$ are
\[
\begin{array}{l}
A = \left(y+\frac{x-y}{p}\right)\cdot\mathbf 1_{p\times p},\\
B = y \cdot\mathbf 1_{p\times q}, \\
C = \left(y+\frac{x-y}{q}\right)\cdot\mathbf 1_{q\times q},\\
\end{array}
\]
and $p= \left\lfloor\frac{n}{2}\right\rfloor$, $q=n-p$.

We also noticed other interesting phenomena, which suggest the need for further geometric analysis.
For example, consider fitting the (non-identifiable) latent class model with $2$ classes to the table of counts (suggested by Bernd Sturmfels)
\[
\left(
\begin{array}{ccc}
5& 1& 1\\
1 & 6 &2\\
1 & 2 &6 \\
\end{array}\right).
\]
Based on our computations, the maximum likelihood estimates appear to be unique, namely the table of fitted values
\begin{equation}\label{eq:othertab}
\left(
\begin{array}{ccc}
5& 1& 1\\
1 & 4 &4\\
1 & 4 &4 \\
\end{array}\right).
\end{equation}

Looking at the non-identifiable subspace for this model, we found that the MLEs (\ref{eq:othertab}) can arise from combinations of parameters some of which can be $0$, such as
\[
\begin{array}{ccc}
\alpha^{(1)} = \beta^{(1)}= \left(
\begin{array}{c}
 0.7143    \\
    0.1429   \\
    0.1429   \\
\end{array}\right),
 & 
 \alpha^{(2)} = \beta^{(2)}= \left(
\begin{array}{c}
0    \\
    0.5   \\
    0.5   \\
\end{array}\right),
& 
\lambda = \left( \begin{array}{c}
   0.3920    \\
    0.6080   \\
\end{array}\right).

\end{array}
\]
This finding seems to indicate the possibility of singularities besides the obvious ones given by marginal probabilities for $H$ containing $0$ coordinates (which have the geometric interpretation as lower order secant varieties) and by points ${\bf p}$ along the boundary of the simplex $\Delta_{d-1}$.

\clearpage

\section{Two Applications}


\subsection{Example: Michigan Influenza}

\cite{MKL:85} present data for 263 individuals on the outbreak of influenza in Tecumseh, Michigan for during the four winters of 1977-1981: (1) Influenza type A (H3N2), December 1977--March 1978; (2) Influenza type A (H1N1), January 1979--March 1979; (3) Influenza type B, January 1980--April 1980 and (4) Influenza type A (H3N2), December 1980--March 1981.    The data have been analyzed by others including  \cite{HABER:86} and we reproduce them here as Table~\ref{tab:mich.flu}.   The table is characterized by a large count for to the cell corresponding to lack of infection from any type of influenza.
 
\begin{table}[ht]
\caption{Infection profiles and frequency of infection for four influenza outbreaks for a sample of 263 individuals in Tecumseh, Michigan during the winters of 1977-1981. A value of of $0$ in the first four columns indicates  Source: ~\cite{MKL:85}. The last column is the values fitted by the naive Bayes model with $r=2$.}
	\centering
	\begin{tabular}{||cccc|r|r||}
		\hline
		\multicolumn{4}{||c}{Type of Influenza} & Observed Counts & Fitted Values \\ \hline
		$(1)$ & $(2)$ & $(3)$ & $(4)$ & & \\ \hline 
		0 & 0 & 0 & 0 & 140 &   139.5135 \\ 
		0 & 0 & 0 & 1 & 31 &     31.3213 \\ 
		0 & 0 & 1 & 0 & 16 &    16.6316 \\ 
		0 & 0 & 1 & 1 & 3&      2.7168 \\ 
		0 & 1 & 0 & 0 & 17&     17.1582 \\ 
		0 & 1 & 0 & 1 & 2 &      2.1122 \\ 
		0 & 1 & 1 & 0 & 5&       5.1172 \\ 
		0 & 1 & 1 & 1 & 1&       0.4292 \\ 
		1 & 0 & 0 & 0 & 20&     20.8160 \\ 
		1 & 0 & 0 & 1 & 2&       1.6975 \\ 
		1 & 0 & 1 & 0 & 9&       7.7354 \\ 
		1 & 0 & 1 & 1 & 0&       0.5679 \\ 
		1 & 1 & 0 & 0 & 12&     11.5472 \\ 
		1 & 1 & 0 & 1 & 1&       0.8341 \\ 
		1 & 1 & 1 & 0 & 4&       4.4809 \\ 
		1 & 1 & 1 & 1 & 0&    0.3209    \\ \hline 
	\end{tabular}
		\label{tab:mich.flu}
\end{table}

The LC model with one binary latent variable \citep[identifiable by Theorem 3.5 in][]{SETSMITH:05} fits the data extremely well, as shown in Table~\ref{tab:mich.flu}. We also conducted a log-linear model analysis of this dataset and concluded that there is no indication of second or higher order interaction among the four types of influenza. The best log-linear model selected via both Pearson's chi-squared and the likelihood ratio statistics was the model of conditional independence of influenza of type (2), (3) and (4) given influenza of type (1) and was outperformed by the LC model. 

Despite the reduced dimensionality of this problem and the large sample size, we report on the instability of the Fisher scoring algorithm implemented in the {\tt R} package {\tt gllm}, e.g., see \cite{ESP:86}. As the algorithm cycles through, the evaluations of Fisher information matrix become increasing ill-conditioned and eventually produce instabilities in the estimated coefficients and in the standard errors. These problems disappear in the modified Newton-Raphson implementation, originally suggested by \cite{HAB:88}, based on an inexact line search method known in the convex optimization literature as the Wolfe conditions.


\subsection{Data From the National Long Term Care Survey}


\cite{EROSHEVA:02} and \cite{EROSHEVA:07} analyze an extract from the National Long Term Care Survey in the form of a $2^{16}$ contingency table  that contains data on $6$ activities of daily living (ADL) and 10 instrumental activities of daily living (IADL) for community-dwelling elderly from 1982, 1984, 1989, and 1994
survey waves. The 6 ADL items include basic activities of hygiene and personal care (eating, getting in/out of bed, getting around inside, dressing, bathing, and getting to the bathroom or using toilet). The
$10$ IADL items include basic activities necessary to reside in the community (doing heavy housework, doing light housework, doing laundry, cooking, grocery shopping, getting about outside, travelling, managing money, taking
medicine, and telephoning). Of the 65,536 cells in the table, 62384 ($95.19\%$) contain zero counts, 1729 ($2.64\%$)contain counts of $1$, 499 ($0.76\%$) contain counts of $2$. The largest cell count, corresponding to the $(1,1,\dots,1)$ cell,  is 3853.

\begin{table}[ht]
\caption{BIC and log-likelihood values for various values of $r$ for the NLTCS dataset.}
\centering
\begin{tabular}{||r|c|c|c||}\hline
$r$ & Dimension & Maximal log-likelihood &  BIC\\ 
\hline
2		&	33	&	-152527.32796 & 305383.97098	\\ 
3		&	50	& -141277.14700 &	283053.25621	\\ 
4		&	67	&	-137464.19759 & 275597.00455	\\ 
5		&	84	&	-135272.97928 & 271384.21508	\\ 
6		&	101	&	-133643.77822 & 268295.46011	\\ 
7		&	118	&	-132659.70775  & 266496.96630	\\ 
8		&	135	&	-131767.71900 & 264882.63595	\\ 
9		&	152	&	-131367.70355  & 264252.25220	\\ 
10	&	169	& -131033.79967 &	263754.09160	\\ 
11	&	186	&	-130835.55275 & 263527.24492	\\ 
12	&	203	&	-130546.33679  & 263118.46015	\\ 
13	&	220	&	-130406.83312 & 263009.09996	 \\ 
14	&	237	&	-130173.98208 & 262713.04502 	\\ 
15	&	254	&	-129953.32247& 262441.37296	\\ 
16	&	271	&	-129858.83550 & 262422.04617	\\ 
17	&	288	& -129721.02032 & 	262316.06296	\\ 
18	&	305	&	-129563.98159 & 262171.63265	\\ 
19	&	322	&	-129475.87848 & 262165.07359	\\ 
20	&	339	&	-129413.69215  & 262210.34807	\\ 
\hline\end{tabular}
\label{tab:2to16.bic}
\end{table}

\cite{EROSHEVA:02} and \cite{EROSHEVA:07} use an individual-level latent mixture model that bears a striking resemblance to the LC model.  Here we report on analyses with the latter.

\begin{table}[!ht]
\caption{Fitted values for the largest six cells for the NLTCS dataset for various $r$.}
\centering
\begin{tabular}{||r|r|r|r|r|r|r||}\hline
$r$ & \multicolumn{6}{c||}{Fitted values} \\\hline
2		&	826.78		&	872.07	&  		6.7&	506.61		&	534.36	 	 &	237.41	  \\	  
3		&	2760.93		&	1395.32	&		152.85	&691.59		&	358.95	 	 &	363.18	 \\
4		&	2839.46	&	1426.07	&		145.13	&	688.54		&	350.58	 	 &	383.19	 \\
5		&	3303.09	&	1436.95	&		341.67	&	422.24		&	240.66	 	 &	337.63	 \\
6		&	3585.98		&	1294.25	&		327.67	& 425.37		&	221.55	 	 &	324.71	 \\
7		&	3659.80		&	1258.53	&		498.76	& 404.57		&	224.22	 	 &	299.52	 \\
8		&	3663.02			&   1226.81			&	497.59	&	411.82			&	227.92			 &	291.99	 \\
9		&	3671.29			&   1221.61			&	526.63	&	395.08			&	236.95			 &	294.54	  \\
10	&	3665.49	&	   1233.16	&		544.95	&	390.92			&	237.69	  		 &	297.72	  \\
11	&	3659.20	&	   1242.27	&		542.72	&	393.12			&	244.37	  		 &	299.26	 \\
12	&	3764.62	&	1161.53	&		615.99	&	384.81			&	235.32	  		 &	260.04	  \\
13	&	3801.73	&	1116.40	&		564.11	&	374.97			&	261.83	  		 &	240.64	  \\
14	&	3796.38	&	1163.62	&		590.33	&	387.73			&	219.89	  		 &	220.34	  \\
15	&	3831.09	&	1135.39	&		660.46	&	361.30			&	261.92	  		 &	210.31	  \\
16	&	3813.80	&	1145.54	&		589.27	&	370.48			&	245.92	  	 &	219.06	 \\
17	&	3816.45	&	1145.45	&		626.85	&	372.89			&	236.16	  		 &	213.25	  \\
18	&	3799.62	&	1164.10	&		641.02	&	387.98			&	219.65	  		 &	221.77	 \\
19	&	3822.68	&	1138.24	&		655.40	&	365.49			&	246.28	  		 &	213.44	  \\
20	&	3836.01	&	1111.51	&		646.39	&	360.52			&	285.27	  		 &	220.47	  \\
\hline
Observed&	3853	&	1107	&	660 & 351 & 303 & 216\\
\hline \hline
\end{tabular}
\label{tab:2to16.fit}
\end{table}

We use both the EM and Newton-Raphson algorithms to fit a number of LC models with up to $20$ classes, which can be shown to be all identifiable  in virtue of Proposition 2.3 in \cite{CGG:02}.  Table \ref{tab:2to16.bic} reports the maximal value of log-likelihood function and the value of BIC (the Bayesian Information Criterion), which seem to indicate that larger LC models with many levels are to be preferred. 
To provide a better sense of how well these LC models fit the data, we show in Table \ref{tab:2to16.fit} the fitted values for the six largest cells, which, as mentioned, deviates considerably from most of the cell entries.   We have also considered alternative model selection criteria such as AIC and modifications of it.  AIC (with and without a 2nd order correction) points to $k>20$! (An ad-hoc modification of AIC due to \cite{anderson} for overdispersed data gives rather bizarre results.)  The dimensionality of a suitable  LC model for these data appears to be much greater than for the individual level mixture model in \cite{EROSHEVA:07}.

Because of its high dimensionality and remarkable degree of sparsity, this example offers an ideal setting in which to test the relative strengths and disadvantages of the EM and Newton-Raphson algorithm.
In general, the EM algorithm, as a hill-climbing method, moves steadily towards solutions with higher value of the log-likelihood, but converges only linearly. On the other hand, despite its faster quadratic rate of convergence, the Newton-Raphson method tends to be very time and space consuming when the number of variables is large, and may be numerically unstable if the Hessian matrices are poorly conditioned around critical points, which again occurs more frequently in large problems (but also in small ones, such as the Michigan Influenza examples above).

For the class of basic LC models considered in this paper, the time complexity for one single step of the EM algorithm is $\mathcal{O} \left( d \cdot r \cdot \sum_i d_i \right)$, while the space complexity is $\mathcal{O} \left( d \cdot r \right)$. In contrast, for the Newton-Raphson algorithm, both the time and space complexity are $\mathcal{O} \left( d \cdot r^2 \cdot \sum_i d_i \right)$. Consequently, for the NLTCS dataset, when $r$ is bigger than 4, Newton-Raphson is sensibly slower than EM, and when $r$ goes up to 7, Newton-Raphson needs more than 1G of memory.
Another significant drawback of the Newton-Raphson method we experienced while fitting both the Michigan influenza and the NLTCS datasets is its potential numerical instability, due to the large condition numbers of the Hessian matrices. As remarked at the end of the previous section, following  \cite{HAB:88}, a numerically convenient solution is to modify the Hessian matrices so that they remain negative definite and then approximate locally the log-likelihood by a quadratic function. However, since the log-likelihood is neither concave and nor quadratic, these modifications do not necessarily guarantee an increase of the log-likelihood at each iteration step. As a result, the algorithm may experience a considerable slowdown  in the rate of convergence, which we in fact observed with the NLTCS data. 
Table \ref{tab:cond.numb} shows the condition numbers for the true Hessian matrices evaluated at the numerical maxima, for various values of $r$. This table suggests that, despite full identifiability, the log-likelihood has a very low curvature around the maxima and that the log-likelihood may, in fact, look quite flat. 

To elucidate this point and some of the many difficulties in fitting LC models, we show in Figure \ref{fig:mesh} the profile likelihood plot for the parameter $\alpha_{12}$ in simplest LC model with $r=2$. The actual profile log-likelihood is shown in red and is obtained as the upper envelop of two distinct, smooth curves, each corresponding to a local maxima of the log-likelihood. The location of the optimal value of $\alpha_{12}$ is displayed with a vertical line. Besides illustrating multimodality,   the log-likelihood function in this example is notable for its relative flatness around its global maximum.


\begin{table}[ht]
\caption{Condition numbers of Hessian matrices at the maxima for the NLTCS data.}
\label{tab:cond.numb}
\centering
\begin{tabular}{|c|c|}\hline
$r$ & Condition number \\\hline
$2$ &      $2.1843e+03$\\
 $3$ &     $1.9758e+04$\\
 $4$ &     $2.1269e+04$\\
 $5$ &     $4.1266e+04$\\
 $6$ &     $1.1720e+08$\\
 $7$ &     $2.1870e+08$\\
 $8$ &     $4.2237e+08$\\
 $9$ &     $8.7595e+08$\\
 $10$ &    $8.5536e+07$\\
 $11$ &    $1.2347e+19$\\
 $12$ &    $3.9824e+08$\\
 $13$ &    $1.0605e+20$\\
 $14$ &    $3.4026e+18$\\
 $15$ &    $3.9783e+20$\\
 $16$ &    $3.2873e+09$\\
 $17$ &    $1.0390e+19$\\
 $18$ &    $2.1018e+09$\\
 $19$ &    $2.0082e+09$\\
 $20$ &    $2.5133e+16$\\\hline
\end{tabular}
\end{table}

\begin{figure}
\centering
\includegraphics[width=5in]{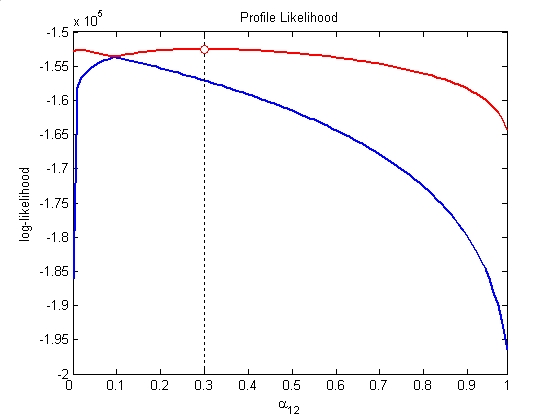}
\caption{The plot of the profile likelihood for the NLCST dataset, as a function of $\alpha_{12}$. The vertical line indicates the location of the maximizer.}
\label{fig:mesh}
\end{figure}

\section{On Symmetric Tables and the MLE}\label{sec:proof}
In this section, we show how symmetry in data allows one to symmetrize  via averaging
local maxima of the likelihood function and to obtain
critical points that are more symmetric.
In various examples we looked at, these have larger 
likelihood than the tables from which they are obtained.
We also prove that if  the aforementioned averaging
process always causes likelihood to go up, then among the $4 \times 4$ matrices of rank 2, the ones maximizing the log-likelihhod function for the 100 Swiss Francs problem (\ref{eq:swiss}) are given in Table \ref{tab:maxima} {\bf a)}. We will further simplify the notation and will write $L$ for the matrix of observed counts and $M$ for the matrix of MLEs.

\subsection{Introduction and Motivation}
A main theme in this section is to understand in what ways symmetry in data  
forces symmetry in the global maxima of the likelihood 
function.  One question
is whether our ideas can be extended at all to 
nonsymmetric data by suitable scaling.
We prove that nonsymmetric local maxima will imply the 
existence of more symmetric points  which are critical
points at least within a key subspace and are related in a very explicit 
way to the nonsymmetric ones.  Thus, if the EM algorithm leads to a local 
maximum which lacks certain symmetries, then one may deduce that certain
other, more symmetric points are also critical points (at least within
certain subspaces), and so check these to see if they give larger 
likelihood.  There is numerical evidence that they do, 
and also a close look at our
proofs shows that for ``many'' data points this symmetrization 
process is guaranteed to increase maximum 
likelihood, by virtue of a certain single-variable
polynomial encoding of the likelihood function 
often being real-rooted.

Here is an example of our symmetrization process.
Given the data 
$$
\begin{array}{cccccc} 
4 & 2 & 2 & 2 & 2 & 2\\
2 & 4 & 2 & 2 & 2 & 2 \\
2 & 2 & 4 & 2 & 2 & 2 \\
2 & 2 & 2 & 4 & 2 & 2 \\
2 & 2 & 2 & 2 & 4 & 2 \\
2 & 2 & 2 & 2 & 2 & 4\\
\end{array},
$$
one of the critical points located by the EM algorithm is
$$
\begin{array}{cccccc} 
7/3 & 7/3 & 7/3 & 7/3 & 7/3 & 7/3\\
7/3 & 13/5 & 13/5 & 13/5 & 29/15 & 29/15 \\
7/3 & 13/5 & 13/5 & 13/5 & 29/15 & 29/15 \\
7/3 & 13/5 & 13/5 & 13/5 & 29/15 & 29/15 \\
7/3 & 29/15 & 29/15 & 29/15 & 44/15 & 44/15 \\
7/3 & 29/15 & 29/15 & 29/15 & 44/15 & 44/15\\
\end{array}.
$$

One way to interpret this matrix is that $M_{i,j} = 7/3 + e_if_j$ 
where 
\[
\bf{e} = \bf{f} 
= (0,2/\sqrt{15},2/\sqrt{15},2/\sqrt{15},
-3/\sqrt{15},-3/\sqrt{15} ).  \]
Our symmetrization process suggests
replacing the vectors $\bf{e}$ and $ \bf{f}$ each by the vector
\[
(1/\sqrt{15}, 1/\sqrt{15} ,2/\sqrt{15} ,2/\sqrt{15} ,
-3/\sqrt{15} ,-3/\sqrt{15})
\] 
in which 
two coordinates are averaged; however, since one of the values being
averaged is zero, it is not so clear whether this should increase
likelihood.  However, repeatedly applying 
such symmetrization steps to this example, 
does converge to a local maximum.  Now let us speak more generally.  
Let $M$ be an $n$ by $n$ matrix of rank at
most two which has row and column sums 
all equalling 
$kn$, implying (by results of Section
~\ref{marginals-section}) that
we may write $M_{i,j}$ as $k + e_if_j$ where $e,f$ are each vectors whose 
coordinates sum to 0.  

We are interested  in the following general question:

\begin{question}\label{smooth-prop}
Suppose a data matrix is fixed under simultaneously
swapping rows and columns $i, j$.  Consider any $M$ as above, i.e.
with $M_{i,j} = k + e_if_j$. 
products also satisfied.  
Does $e_i > e_j > 0, f_i > f_j > 0 $ (or similarly 
$e_i < e_j < 0, f_i< f_j < 0$ )  imply that replacing 
$e_i,e_j$ each by $\frac{e_i + e_j}{2}$ and $f_i,f_j$
each by $\frac{f_i+f_j}{2}$  always 
increases  the likelihood?
\end{question}

\begin{remark}
The weaker conditions $e_i > e_j = 0$ and
$f_i > f_j = 0$ (resp. $e_i < e_j = 0, f_i < f_j = 0$) 
do not always 
imply that this  replacement will increase 
likelihood.   However, one may consider the finite list of 
possibilities for how many zeroes the vectors $\bf{e}$ and $\bf{f}$ may each
have; an affirmative answer to Question ~\ref{smooth-prop} would give
a way to find the matrix maximizing likelihood in each case, and then
we could compare this finite list of maxima to find the global 
maximum.
\end{remark}

\begin{question}
Are all real-valued critical points of the likelihood function obtained
by setting some number of coordinates in the $\bf{e}$ and $\bf{f}$ 
vectors to zero
and then averaging by the above process so that the eventual vectors 
$\bf{e}$ and $\bf{f}$ have all positive 
coordinates equal to each other and all negative
coordinates equal to each other?  This seems to be true in many 
examples.
\end{question}

One may check that 
the example discussed in Chapter 1 of \cite{PS:05}  
gives another instance where this averaging
approach leads quickly to what appears to be
a global maximum.  
Namely, given the data matrix 
$$
\begin{array}{cccc}
4 & 2 & 2 & 2\\
2 & 4 & 2 & 2\\
2 & 2 & 4 & 2\\
2 & 2 & 2 & 4\\
\end{array}$$
and a particular starting point, 
the EM algorithm converges to the saddle point
$$
\frac{1}{48}
\begin{array}{cccc}
4 & 2 & 3 & 3\\
2 & 4 & 3 & 3\\
3 & 3 & 3 & 3\\
3 & 3 & 3 & 3\\
\end{array},$$
whose entries may be written as 
$M_{i,j} = 1/48 (3 + a_ib_j)$ for $\bf{a} = (-1,1,0,0)$ and $\bf{b} 
= (-1,1,0,0)$.  Averaging $-1$ with 
$0$ and $1$ with the other $0$ simultaneously in $\bf{a}$ and $\bf{b}$ 
immediately yields the
global maximum directly by symmetrizing the saddle point, i.e. rather 
than finding it
by running the  EM algorithm repeatedly from various starting points.  

An affirmative answer to Question
~\ref{smooth-prop} would imply several things.  
It would yield a (positive) solution to the 100
Swiss francs problem, as discussed in Section ~\ref{swiss-section}.  
More generally, it 
would explain in a rather precise way how certain
symmetries in data seem to impose symmetry on the global maxima of the maximum
likelihood function.  Moreover it would suggest good ways to look for
global maxima, as well as constraining them enough that in 
some cases they can be
characterized, as we demonstrate for the 100 Swiss francs problem.   To make
this concrete, one thing it would tell us for an $n$ by $n$ data matrix which
is fixed by the $S_n$ action simultaneously permuting rows and columns in the 
same way, is that any probability matrix maximizing likelihood for such a 
data matrix will have at most two distinct types of rows. 

We do not know the answer to this question, but we do prove that this type
of averaging will at least give a critical point within the subspace in which
$e_i,e_j,f_i,f_j$ may vary freely but all other parameters are held fixed.
Data also provides evidence that the answer to the question may very well
be yes.  At the very least, this 
type of averaging appears to be a good heuristic for seeking local
maxima, or at least finding a way to continue to increase maximum
likelihood beyond what it is at a critical point one reaches.  
Moreover, while real data is unlikely to have these symmetries, perhaps
it could come close, and this could still be a good heuristic to use 
in conjunction with the EM algorithm.


\subsection{Preservation of Marginals and Some 
Consequences}\label{marginals-section}

\begin{proposition}\label{row-sum-equal}
Given data in which all row and column sums (i.e. marginals) are equal, then 
for $M$ to maximize the likelihood function for this data among matrices
of a fixed rank, 
row and column sums of $M$ all must be equal.
\end{proposition}

We prove the case mentioned in
the abstract, which should generalize by adjusting
exponents and ratios in the proof.  It may very well also 
generalize to distinct
marginals and  tables with more rows
and columns.

\begin{proof}
Let $R_1,R_2,R_3,R_4$ be the row sums of $M$.  Suppose
$R_1\ge R_2 \ge R_3 > R_4$; other cases will be similar.  Choose $\delta $ 
so that $R_3 = (1 + \delta )R_4$.  We will show that multiplying row 4 by
any $1 + \epsilon $ with $0 < \epsilon < \min (1/4, \delta /2 )$ will strictly
increase $L$, giving a contradiction to $M$ 
maximizing $L$.  The result for column sums follows by symmetry.

Let us write $L(M')$ for the new matrix $M'$
in terms of the variables $x_{i,j}$ for
the original matrix $M$, so as to show that $L(M') > L(M)$.
The first inequality below is proven in Lemma ~\ref{row-lemma}.

\begin{eqnarray*}
L(M') &=& \frac{(1 + \epsilon)^{10} (\prod_{i=1}^4 x_{i,i})^4 (\prod_{i\ne j}
x_{i,j})^2 }{R_1 + R_2 + R_3 + (1+\epsilon )R_4 )^{40}}\\
  &> & \frac{(1 + \epsilon )^{10} (\prod_{i=1}^4 x_{i,i})^4 (\prod_{i\ne j}
x_{i,j})^2 }{[(1+1/4(\epsilon - \epsilon^2))(R_1 + R_2 + R_3 + R_4)]^{40}}\\
&=&\frac{(1+\epsilon)^{10} (\prod_{i=1}^4 x_{i,i})^4 (\prod_{i\ne j}x_{i,j})^2}{[(1+ 1/4(\epsilon -\epsilon^2))^4]^{10}[R_1 + R_2 + R_3 + R_4]^{40}}\\
&=&\frac{(1+\epsilon)^{10} (\prod_{i=1}^4 x_{i,i})^4 (\prod_{i\ne j}x_{i,j})^2}{[1 + 4(1/4)(\epsilon - \epsilon^2 ) + 6(1/4)^2(\epsilon - \epsilon^2)^2 + \cdots
+ (1/4)^4(\epsilon - \epsilon^2)^4 ]^{10} [\sum_{i=1}^4R_i]^{40}} \\
&\ge & \frac{(1+\epsilon)^{10}}{(1+\epsilon)^{10}}\cdot L(M)\\
\end{eqnarray*}
\end{proof}

\begin{lemma}\label{row-lemma}
If $\epsilon < \min (1/4,\delta /2)$ and $R_1 \ge R_2 \ge R_3 = (1+\delta )R_4$, then $R_1 + R_2 + R_3 + (1+\epsilon )R_4 < (1 + 1/4(\epsilon - \epsilon^2))(R_1 + R_2 + R_3 + R_4 )$.
\end{lemma}

\begin{proof}
It is equivalent to show $\epsilon R_4 < (1/4)(\epsilon)(1-\epsilon)\sum_{i=1}^4
R_i$.  However,
\begin{eqnarray*}
(1/4)(\epsilon)(1-\epsilon )(\sum_{i=1}^4 R_i ) &\ge& (3/4)(\epsilon )(1-
\epsilon )(1 + \delta )R_4 + (1/4)(\epsilon )(1-\epsilon )R_4 \\
&>& (3/4)(\epsilon )(1- \epsilon )(1 + 2\epsilon )R_4 + (1/4)(\epsilon )(1 - \epsilon )R_4 \\
&=& (3/4)(\epsilon )(1 + \epsilon - 2\epsilon^2 )R_4 + (1/4)(\epsilon -\epsilon^2)R_4 \\
&=& \epsilon R_4 + [(3/4)(\epsilon^2) - (6/4)(\epsilon^3) ]R_4 - (1/4)(\epsilon^2 )R_4 \\
&=& \epsilon R_4 + [(1/2)(\epsilon^2) - (3/2)(\epsilon^3)]R_4 \\
&\ge & \epsilon R_4 + [(1/2)(\epsilon^2) - (3/2)(\epsilon^2)(1/4)]R_4 \\
&>& \epsilon R_4 .\\
\end{eqnarray*} 
\end{proof}

\begin{corollary}\label{parameter-cor}
There exist vectors $(e_1,e_2,e_3,e_4)$ and $(f_1,f_2,f_3,f_4)$ such
that $\sum_{i=1}^4 e_i = \sum_{i=1}^4 f_i = 0$ and 
$M_{i,j} = K + e_if_j$.
Moreover, $K$ equals the average entry size.
\end{corollary}

In particular, this tells us
that $L$ may be maximized by treating it as a function of just six 
variables, namely 
$e_1,e_2,e_3,f_1,f_2,f_3$, since $e_4,f_4$ are also determined by
these; changing $K$ before solving this maximization problem simply has
the impact of 
multiplying the entire matrix $M$ that maximizes likelihood by a scalar.

Let $E$ be the {\it deviation matrix} associated to $M$, where $E_{i,j}
= e_if_j$.

\begin{question}
Another natural question to ask, in light of this corollary, is whether
the matrix of rank at most $r$ maximizing $L$ is expressible as the sum of a 
rank one matrix and a matrix of rank at most $r-1$ that  maximizes
$L$ among matrices of rank at most $r-1$.
\end{question}

\begin{remark}
When we consider matrices with fixed row and column sums, then we
may ignore the denominator in the likelihood function and
simply maximize the numerator.
\end{remark}

\begin{corollary}
If $M$ which maximizes $L$ has $e_i=e_j$, then it also has $f_i=f_j$.
Consequently, if it has $e_i \ne e_j$, then it also has $f_i\ne f_j$.
\end{corollary}

\begin{proof}
One consequence of having equal row and column sums is that it allows 
the likelihood function to be split into a product of four functions,
one for each row, or else one for each column; this is because the 
sum of all table entries equals the sum of those in any row or column 
multiplied by four, allowing the denominator to be written just using 
variables from any one row or column.  Thus, once the vector $e$ is
chosen, we find the best possible $f$ for this given $e$ by solving
four separate maximization problems, one for each $f_i$, i.e. one for
each column.  Setting $e_i=e_j$ causes the likelihood function for 
column $i$ to coincide with the likelihood function for column $j$, so
both are maximized at the same value, implying $f_i=f_j$.
\end{proof}

Next we prove a slightly stronger general fact for matrices in which 
rows and columns $i,j$ may simultaneously be swapped without changing
the data matrix:

\begin{proposition}\label{same-rel-order}
If a matrix $M$ maximizing likelihood 
has $e_i > e_j>0$, then it also has $f_i > f_j>0$.
\end{proposition}

\begin{proof}
Without loss of generality, say $i=1,j=3$.  We will show that if
$e_1>e_3$ and $f_1<f_3$, then swapping columns one and three will 
increase likelihood, yielding a contradiction.  
 Let
$$L_1(e_1) = (1/4 + e_1f_1)^4(1/4 + e_1f_2)^2(1/4 + e_1f_3)^2(1/4 + e_1f_4)^2$$
and
$$L_3(e_3) = (1/4 + e_2f_1)^2 (1/4 + e_2f_2)^2(1/4 + e_3f_3)^4 (1/4 + 
e_3f_4)^2,$$
namely the contributions of rows 1 and 3 to the likelihood function.
Let
$$K_1(e_1) = (1/4 + e_1f_3)^4(1/4 + e_1f_2)^2(1/4 + e_1f_1)^2(1/4 + 
e_1f_4)^2$$
and
$$K_3(e_3) = (1/4 + e_3f_3)^2 (1/4 + e_3f_2)^2(1/4 + e_3f_1)^4 (1/4 + 
e_3f_4)^2,$$
so that after swapping the first and third columns, the new contribution 
to the likelihood function from rows one and three is
$K_1(e_1)K_3(e_3)$.
Since the column swap does not 
impact that contributions from rows 2 and 4,
the point is to show $K_1(e_1)K_3(e_3) > L_1(e_1)L_3(e_3)$. 
Ignoring 
common factors, this reduces to showing
$$(1/4 + e_1f_3)^2(1/4 +e_3f_1)^2 > (1/4 + e_1f_1)^2 
(1/4 + e_3f_3)^2,$$ in other
words
$$(1/16 + 1/4(e_1f_3 + e_3f_1) + e_1e_3f_1f_3)^2 > (1/16 + 1/4(e_1f_1 
+ e_3f_3) + e_1e_3f_1f_3)^2,$$ 
namely $e_1f_3 + e_3f_1 > e_1f_1 + e_3f_3$.  But since
$e_3<e_1,f_1<f_3$, we have $0 < (e_1-e_3)(f_3-f_1) = (e_1f_3 + 
e_3f_1) - (e_1f_1 + e_3f_3)$, just as needed. 
\end{proof}

\begin{question}
Does having a data matrix which is symmetric with respect to transpose
imply that matrices maximizing likelihood will also
be symmetric with respect to transpose?  
\end{question}

Perhaps this could also be verified again by averaging, similarly to 
what we suggest for involutions swapping a pair of rows and columns
simultaneously.

\subsection{The 100 Swiss Francs Problem}\label{swiss-section}

We use the results derived to far so show how to reduce the 100 Swiss Francs problem to
Question ~\ref{smooth-prop}.  Thus, an affirmative answer to Question ~\ref{smooth-prop}
would  provide a mathematical proof formally that the
three tables in \ref{tab:maxima} {\bf a)} are global maxima of the log-likelihood function for the basic LC model with $r=2$ and data given in (\ref{eq:swiss}).

\begin{theorem}
If the answer to Question ~\ref{smooth-prop} is yes, then the 100 Swiss francs problem is solved.
\end{theorem}

\begin{proof}
Proposition ~\ref{row-sum-equal} 
showed that for $M$ to maximize $L$, $M$ must have row and 
column sums which are all equal to the quantity which we call $R_1,R_2,R_3,
R_4,C_1,C_2,C_3,$ or $C_4$ at our convenience.  
The denominator of $L$ may therefore
be expressed
as $(4C_1)^{10}(4C_2)^{10}(4C_3)^{10}(4C_4)^{10}$ or as
$(4R_1)^{10}(4R_2)^{10}(4R_3)^{10}(4R_4)^{10}$, enabling us to rewrite $L$
as a product of four smaller functions using distinct sets of variables.

Note that letting $S_4$ simultaneously 
permute rows and columns will not change $L$, so let us assume the first 
two rows of $M$ are linearly independent.  Moreover, we may choose the first
two rows in such a way that the 
next two rows are each nonnegative combinations of the first two.  Since 
row and column sums are all equal, the third row, denoted $v_3$, is 
expressible as $xv_1 + (1-x)v_2$ for $v_1,v_2$ the first and second rows and
$x \in [0,1]$.    One may check that $M$ does not have any row or column 
with values all equal to each other, because if it had one, then it would have the
other, reducing to a three by three problem which one may solve, and one
may check that the answer does not have as high of likelihood as 
$$\begin{array}{cccc}
3 & 3 & 2 & 2\\
3 & 3 & 2 & 2\\
2 & 2 & 3 & 3\\
2 & 2 & 3 & 3\\
\end{array}.
$$
Proposition ~\ref{repeat-prop} will show that 
if the answer to Question ~\ref{smooth-prop} is yes, then 
for $M$ to maximize $L$, we must have 
$x=0$ or $x=1$, implying row 3 equals either row 1 or row 2, and likewise
row 4 equals one of the first two rows.  Proposition 
~\ref{no-three-prop}
shows $M$ does not have three rows all equal to each other, and therefore
must have two pairs of equal rows.  Thus, the first column takes the form
$(a,a,b,b)^T$, so it is simply a matter of optimizing $a$ and $b$, then 
noting that the optimal choice 
will likewise optimize the other columns (by virtue of the way we broke 
$L$ into a product of four expressions which are essentially the same, one
for each column).  Thus, $M$ takes the form
$$
\begin{array}{cccc}
a & a & b & b \\
a & a & b & b \\
b & b & a & a \\
b & b & a & a \\
\end{array}
$$
since this matrix does indeed have rank two.  Proposition 
~\ref{one-d-prop} shows that to
maximize $L$ one needs $2a = 3b$, finishing the proof.
\end{proof}

\begin{proposition}\label{repeat-prop}
If the answer to Question ~\ref{smooth-prop} is yes, then 
row 3 equals either row 1 or row 2 in any matrix $M$ which maximizes
likelihood. Similarly, each row $i$
with $i>2$ equals either row 1 or row 2.
\end{proposition}

\begin{proof}
$M_{3,3} = xM_{1,3} + (1-x)M_{2,3}$ for some $x\in [0,1]$, so
$M_{3,3} \le \max (M_{1,3},M_{2,3} )$.  
If $M_{1,3} = M_{2,3}$, then all entries of this column are equal, and
one may use calculus to eliminate this possibility as follows: either
$M$ has rank one, and then we may replace column three by $(c,c,2c,c)^T$
for suitable constant $c$ to increase likelihood, since this only increases
rank to at most two, or else the column space of $M$ is spanned by 
$(1,1,1,1)^T$ and some $(a_1,a_2,a_3,a_4)$ with $\sum a_i = 0$; specifically,
column three equals $(1/4,1/4,1/4,1/4) + x(a_1,a_2,a_3,a_4)$ for some $x$,
allowing its contribution to the likelihood function to be expressed as a 
function of $x$ whose derivative at $x=0$ is nonzero, provided that
$a_3\ne 0$, implying that adding
or subtracting some small multiple of $(a_1,a_2,a_3,a_4)^T$ to the column
will make the likelihood increase.  If $a_3=0$, then row three is also
constant, i.e. $e_3 = f_3 = 0$.  But then, an affirmative answer to the 
second part of Question ~\ref{smooth-prop} will imply that this matrix
does not maximize likelihood.


Suppose, on the other hand, $M_{1,3}> M_{2,3}$. Our goal then is to 
show $x=1$.  By Proposition
~\ref{row-sum-equal} applied to columns rather than rows, 
we know that $(1,1,1,1)$ is in the span of the 
rows, so each row may be written as $1/4 (1,1,1,1) + cv$ for some fixed
vector $v$ whose coordinates sum to 0.  Say
row 1 equals $1/4 (1,1,1,1) + kv$ for $k=1$.  Writing row three as
$1/4 (1,1,1,1) + lv$, what remains is to 
rule out the possibility $l<k$.
However, Proposition ~\ref{same-rel-order} shows that 
$l<k$ and $a_1<a_3$ together imply that 
swapping columns one and three will yield a new matrix of
the same rank with larger likelihood. 

Now we turn to the case of 
$l<k$ and $a_1\ge a_3$.  If $a_1 = a_3$ then
swapping rows one and three will increase likelihood. 
Assume $a_1 > a_3$.  By Corollary
~\ref{parameter-cor}, we have $(e_1,e_2,e_3,e_4)$ with $e_1 > e_3 $ and 
$(f_1,f_2,f_3,f_4) $ with $f_1 > f_3 $.  Therefore, if the answer to
Question ~\ref{smooth-prop} is yes, then
replacing $e_1,e_3$ each by $\frac{e_1 + e_3}{2} $ and
$f_1,f_3$ each by $\frac{f_1+f_3}{2}$ yields a matrix with larger
likelihood, completing the proof.
\end{proof}

\begin{proposition}\label{no-three-prop}
In any matrix $M$ maximizing $L$ among rank 2 matrices, no three rows of
$M$ are equal to each other.
\end{proposition}

\begin{proof}
Without loss of generality,
if $M$ had three equal rows, then $M$ would take the form
$$
\begin{array}{cccc}
a & c & e & g\\
b & d & f & h\\
b & d & f & h\\
b & d & f & h\\
\end{array}
$$
but then the fact that $M$ maximizes $L$ ensures $d=f=h$ and $c=e=g$
since $L$ is a product of four expressions, one for
each column, so that the second, third and fourth columns will all  
maximize their contribution to $L$ in the same way.  Since all row and 
column sums are equal, simple algebra may be used to show that all
entries must be equal.  However, 
we have already shown that such matrices do not 
maximize $L$. 
\end{proof}

\begin{proposition}\label{one-d-prop}
To maximize $M$ requires $a,b$ related by $2a = 3b$.
\end{proposition}

\begin{proof}
We must maximize $\frac{a^6b^4}{(8a+8b)^{10}}$.  We may assume $a + b =1$ 
since multiplying the entire matrix by a constant does not change $L$, so
we maximize $(1/8)^{10}a^6b^4$ 
with $b=1-a$; in other words, we 
maximize $f(a) = a^6(1 - a)^4$.  But solving $f'(a) = 0 = 6a^5 (1 - a)^4
+ a^6(4)(1 -a)^3(-1)  = a^5(1-a)^3[6(1-a) -4a]$ yields
$6(1-a)-4a=0$, so $a=6/10$ and $b=4/10$ as desired.
\end{proof}

\section{Conclusions}

In this paper we have reconsidered the classical latent class model for contingency table data and studied its geometric and statistical properties.  For the former we have exploited tools from algebraic geometry and computation tools that have allowed us to display the complexities of the latent class model.  We have focused on the problem of maximum likelihood estimation under LC models and have studied the singularities arising from symmetries in the contingency table data and the multiple maxima that appear to result from these.  We have given an informal characterization of this problem, but a strict mathematical proof of the existence of identical multiple maxima has eluded us; we describe elements of a proof in a separate section.

We have also applied LC models to data arising in two real-life applications.  
In one, the model is quite simple and maximum likelihood estimation poses little problems, whereas in the other high-dimensional example various issues, computational as well as model-based, arise.  From computational standpoint, both the EM and the Newton-Raphson algorithm are especially vulnerable to problems of multimodality and provide little in the way of clues regarding the dimensionality difficulties associated with the underlying structure of LC models.  Furthermore, the seemingly singular behavior of the Fisher information matrix at the MLE that we observe even for well-behaved, identifiable models is an additional element of complexity.


Based on our work, we would advise practitioners to exercise caution in applying LC models, especially to sparse  data.  They have a tremendous heuristic appeal and in some examples provide a clear and convincing description of the data.  But in many situations, the kind of complex behavior explored in this paper may lead to erroneous inferences.

\section{Acknowledgments} 
This research was supported in part by the National Institutes of Health under Grant No. R01 AG023141-01, by NSF Grant DMS-0631589, and by a grant from the Pennsylvania Department of Health through the Commonwealth Universal Research Enhancement Program, all to the Department of Statistics to Carnegie Mellon University,  and  by NSF Grant DMS-0439734 to the Institute for Mathematics and Its Application at the University of Minnesota.  We thank Bernd Sturmfels for introducing us to the 100 Swiss Francs problem,  which motivated much of this work, and for his valuable comments and feedback.

\end{document}